\documentclass[12pt]{iopart}

\usepackage[T1]{fontenc}
\usepackage{iopams}
\expandafter\let\csname equation*\endcsname\relax
\expandafter\let\csname endequation*\endcsname\relax
\usepackage{amsmath}
\usepackage{amsfonts}
\usepackage{amssymb}
\usepackage{amsthm}
\usepackage{color}
\newcommand{\lt}{\left }
\newcommand{\rt}{\right}
\newcommand{\R}{\mathbb{R}}
\newcommand{\Ref}[1]{(\ref{#1})}
\newcommand{\bgeq}{\begin{equation}}
\newcommand{\eeq}{\end{equation}}
\newcommand{\bg}{\begin}
\newcommand{\bd}[1]{\boldsymbol{#1}}
\newcommand{\I}{\mathrm{i}}
\newcommand{\deq}{\overset{d}{=}}
\newcommand{\defeq}{\mathrel{\mathop:}=}
\newcommand{\dd}{\mathrm{d}}

\newcommand{\f}[2]{\frac{#1}{#2}}

\newcommand{\const}{\mathrm{const.}}
\newcommand{\EE}[1]{\E\lt[ #1 \rt]}
\newcommand{\E}{\mathbb{E}}
\usepackage{graphicx}

\bg{document}

\title{Codifference can detect ergodicity breaking and non-Gaussianity}

\author{Jakub {\'S}l\k{e}zak$^{\dagger\ddag}$, Ralf Metzler$^\sharp$, and Marcin
Magdziarz$^\ddag$}
\address{$^\dagger$Department of Physics, Bar Ilan University\\
$^\ddag$Faculty of Pure and Applied Mathematics, Wroc{\l}aw University of
Science and Technology\\
$^\sharp$Institute of Physics and Astronomy, Potsdam University}
\ead{rmetzler@uni-potsdam.de. Corresponding author: Ralf Metzler}

\bg{abstract}
We show that the codifference is a useful tool in studying the ergodicity breaking
and non-Gaussianity properties of stochastic time series. While the codifference is
a measure of dependence that was previously studied mainly in the context of stable
processes, we here extend its range of applicability to random-parameter and
diffusing-diffusivity models which are important in contemporary physics, biology
and financial engineering. We prove that the codifference detects forms of dependence
and ergodicity breaking which are not visible from analysing the covariance and
correlation functions. We also discuss a related measure of dispersion, which is a
non-linear analogue of the mean squared displacement.
\end{abstract}

\setcounter{equation}{0}

\section{Introduction}
\subsection{Statistical measures in modelling of diffusion}

The analysis of stochastic systems has three important and partially distinct
aspects: models, properties and estimation. These roughly correspond to
physical, mathematical and statistical aspects of research. Modelling is
concerned with explaining the nature of a  system according to the underlying
theory (e.g. "the particle undergoes Brownian motion, because it rapidly
exchanges momenta with the molecules of liquid"). The analysis of statistical
properties (also called "measures")\footnote{Strictly speaking, these
are sufficiently regular functionals acting on the space of observations. In
quantum mechanics each such linear functional corresponds to an observable. In
statistical mechanics a similar r{\^o}le is fulfilled by $\E[f(X)]$ for bounded
continuous functions $f$. In statistics linearity is usually not required and
various measures have the form $g(\E[f(X)])$. This is the case also in the
present work.}
relates these models with observable quantities ("Brownian motion has a linear
mean squared displacement"). By using suitable estimators we link these
parameters to the experimental data ("the mean squared displacement can be
efficiently estimated by an arithmetic average over squared displacements").

This work is motivated by our conviction that the choice of statistical
measures is too small for contemporary needs, as the scope and number of models
increased considerably \cite{pccp}. The classical models based on the Langevin
equation \cite{coffey}, the generalised Langevin equation \cite{lutz,kou}, as;
well as short- \cite{Codling} and long- \cite{restaurant,Schulz18}
memory random walks were complemented by motions on fractals \cite{diffFrac},
motions in complex energy landscapes \cite{Camacho}, random walks in random
environments \cite{exSinaiDiff,bouchoud}, random walks with correlated steps
and waiting times
\cite{Renshaw,Bovet,vincent,marcin,johannes} and L{\'e}vy walks \cite{Zaburdaev},
spatially heterogeneous diffusion processes \cite{ergPar}, diffusing-diffusivity
\cite{diffDiff} and more. Distinguishing between different models from this
wide class is of course crucially dependent on the physical understanding of the
system, but this requirement does not lessen the  importance of empirical
verification based on various measures and corresponding estimators. From an
experimental point of view the large range of different stochastic processes
is called for by ever more detailed insights garnered in highly complex
environments such as living biological cells or membranes, for instance, by
single particle tracking of individual sub-micron tracers of even fluorescently
labelled single molecules \cite{hoefling,norregaard,bbm}.

Traditionally, in the study of diffusion phenomena, the three most basic and
popular statistical measures in use are: the mean as a measure of location,
the mean squared displacement (MSD) as a measure of dispersion and the
covariance as a measure of dependence, respectively
\bgeq
\mu_X(t)\defeq \E[X_t],\quad \delta^2_X(t)\defeq \E[ (X_t-\mu_X(t))^2],\quad
r_X(t)\defeq \E[ (X_{s+t}-\mu_X(s+t))(X_s-\mu_X(s))].
\eeq
Other, alternative choices of measures could be, for example: the
median for the location \cite{statDict}, entropy \cite{entropy} or
quantile ranges \cite{statDict} for the dispersion, the rank correlation
\cite{statDict,kendall} or the mutual information \cite{mutualInfo} for
the dependence.

The covariance as defined above should not depend on the choice of $s$,
which is true for stationary processes (the term "non-ageing" is also in
use). We will assume stationarity whenever we will be studying memory. In
practical applications this condition is fulfilled by many types of confined
motions or increments of free diffusions.  The more general non-stationary
case will be only briefly mentioned in Eqs. \Ref{eq:covDefFull} and
\Ref{eq:codDefFull}. Many of the arguments presented here could be further
extended to non-stationary models, but it would require a case-by-case
study. Conversely, measures of dispersion and location are interesting
mostly for non-stationary (ageing) processes, otherwise they are constant,
and the discussed cases will fit into that category.

The present range of typically employed measures, which could be effectively
used for studying diffusion is indeed quite limited, and the need of a wider range
of methods has been acknowledged for many years. Various papers proposed,
e.g. studying  higher order moments and ratios of moments \cite{momentRatios},
running maximum \cite{tejedor}, p-variation \cite{pvar}, or time averages
and ensemble averages of time averages \cite{neuralSurf}. A prominent
example of the last kind of measure is, e.g., the ergodicity breaking parameter
\cite{ergPar,ergBreakM,He08,weakErgJeon}. Recently also single-trajectory power
spectral methods were proposed \cite{gleb,gleb1}. These techniques are steadily
gaining public recognition, but often the range of their application
is still narrow. Moreover, a large part of this important research has a
limitation of studying properties "not very different" from the second order
on. For example, any power function $x^\alpha$ for $\alpha>1$ has a similar
behaviour to $x^2$ (i.e., it is an increasing, convex function) and parameters
based on it are usually not far away from the classical ones.\footnote{This
similarity is what causes the "strong anomalous diffusion" property, for
which the power-law dependency $\E|X_t|^q\propto t^{q v(q)}$ is observed for
non-constant function $v$ \cite{strongAn}.} They all emphasise highly the
tails of the distribution, and any change of distributions for large values of
observations has a larger influence than for the small ones. This connection
is very helpful in making comparisons, but the important part of the total
information is lost and could be extracted using more distinct measures.

\subsection{Overview of the codifference}

Our main subject of interest, the codifference, is an example for a measure
different from those based on moments. It was initially proposed as a tool
to measure the dependence for $\alpha$-stable processes, for which the second
moment is infinite \cite{taqqu,kokoszka,kokoszka2, stableOU, stableFOU,
solarFlare}.  However, in many systems the divergence of the second
moment is not an expected physical property, which limits the range of
possible applications of stable processes. It was already noticed, e.g., in
\cite{codWol}  that the codifference may be useful for both models with or
without finite second moment. In our present work we study the applications
of the codifference for a class of models based on Gaussian distributions,
which we call  conditionally Gaussian processes; as we will demonstrate many
useful and widely used models fit into this category.

The definition of the codifference which we will use is as follows: for any stationary
process $X$ it is given by the formula
\bgeq\label{eq:cdfDef}
\tau_X^\theta(t)\defeq \f{1}{\theta^2}\ln\f{\E\lt[\e^{\I\theta(X_{s+t}-X_s)}\rt]}{\E\lt[\e^{\I\theta X_{s+t}}\rt]\E\lt[\e^{-\I\theta X_s}\rt]}.
\eeq
The sample codifference is introduced in a standard way, by replacing the three
ensemble averages $\E[\bd\cdot]$ in the above expression by arithmetic averages
$\f{1}{n}\sum_{j=1}^n (\bd\cdot)$. Similarly, one can consider a time-averaged
codifference. For all symmetric distributions the considered averages should
be real-valued, so in most of the practical applications one can average
over $\cos(\theta(\bd\cdot))$ instead of $\exp(\I\theta(\bd\cdot))$; this
was used for the Monte Carlo simulations which will be presented further on.

Note that the so-called generalised codifference has $X_{s+t}$ and $X_s$
multiplied by $\theta_1$ and $\theta_2$ respectively and contains even more
information \cite{taqqu}. In the context of models that we will consider
this additional flexibility does not seem to be meaningful and so the cost
of complicating our formulae would be unreasonable.

Conversely, the basic formula for the codifference in the classical
book of Samorodnitsky and Taqqu \cite{taqqu} is similar to ours, but
with $\theta=1$. In the mathematical study of stable process this is
sufficient, but in more broad physical applications introducing an arbitrary
dimensional constant equal to unity is not desirable. In our choice of
definition the codifference has the unit of $X^2$ due to the introduction of
$1/\theta^2$. This factor makes the codifference comparable to the covariance,
and allows us to show them on the same plots.  When this is not important
the factor $1/\theta^2$ can be omitted. There exists an even more simplified
object, the dynamical functional \cite{dynFunc}, which is just the numerator
minus the denominator from \Ref{eq:cdfDef} with $\theta = 1$; it is used to
study ergodicity breaking \cite{neuralSurf, epsErgBreak}.

Instead of moments such as the covariance, the codifference depends on sines and
cosines of $\theta X_{s+t}$ and $\theta X_s$. Expanding these functions into
Taylor series around zero up to the two first terms and using the fact that
for stationary process $\EE{X_t}=\const$  shows that the codifference agrees
with the covariance for distributions concentrated around the origin. The
most essential difference is that the codifference measures mainly the
dependence determined by the bulk of the probability density in contrast
to the covariance, which puts much larger emphasis on the tails. This is
caused by the cancellation of highly oscillatory terms in the tails of the
PDF as stated by the Riemann-Lebesgue lemma, which is in contrast to the huge
influence of the tails in the covariance caused by the quadratic factor in
the probabilistic integral $\E[X_s X_{s+t}]$.

Because of the presence of two highly non-linear transformations:
sine/cosine and logarithm, definition \Ref{eq:cdfDef} may initially
not seem very intuitive. It becomes more natural if we interpret it as a
conveniently transformed Fourier transform of the distribution (that is, the
probabilistic characteristic function). In the full, multidimensional form, the
characteristic function contains all information about the dependence. Moreover
for Gaussian variables it has the very simple form $\exp(-(\theta\sigma)^2/2)$,
so it seems reasonable to use it as a dependence measure  for models related
to the Gaussian distribution. Still, it is not obvious that the codifference
behaves as we would require from a memory function. Fortunately, simple
arguments show that this is the case:

\bg{itemize}

\item[a)] When $X_{s+t}=X_s$ (the case of total positive dependence) the codifference is a positive constant $\tau_X^\theta(t)=\tau_X^\theta(0)>0$. If the values $X_{s+t}$ and $X_s$ become independent, the codifference converges to 0. Both facts are immediate consequences of the definition together with 
\bgeq
\E\lt[\e^{\I\theta X_{s+t}}\rt]\E\lt[\e^{-\I\theta X_s}\rt] = \lt|\E\lt[\e^{\I\theta X_s}\rt]\rt|^2<1
\eeq
and, for $X_{s+t}$ independent of $X_s$,
\bgeq
\E\lt[\e^{\I\theta(X_{s+t}-X_s)}\rt] = \E\lt[\e^{\I\theta X_{s+t}}\rt]\E\lt[\e^{-\I\theta X_s}\rt].
\eeq

\item[b)]If the process is a sum of independent components $X_t=Y_t+Z_t$ then the respective codifferences are additive
\bg{align}
\tau_X^\theta(t)&=\ln\f{\E\lt[\e^{\I\theta(Y_{s+t}-Y_s)}\e^{\I\theta(Z_{s+t}-Z_s)}\rt]}{\E\lt[\e^{\I\theta Y_{s+t}}\e^{\I\theta Z_{s+t}}\rt]\E\lt[\e^{-\I\theta Y_s}\e^{-\I\theta Z_s}\rt]} = \ln\f{\E\lt[\e^{\I\theta(Y_{s+t}-Y_s)}\rt]\E\lt[\e^{\I\theta(Z_{s+t}-Z_s)}\rt]}{\E\lt[\e^{\I\theta Y_{s+t}}\rt]\E\lt[\e^{\I\theta Z_{s+t}}\rt]\E\lt[\e^{-\I\theta Y_s}\rt]\E\lt[\e^{-\I\theta Z_s}\rt]}\nonumber\\
&=\tau_Y^\theta(t)+\tau_Z^\theta(t).
\end{align}
This property is important in common applications, where the observed process
usually is at least to some degree disturbed by noise, which can most often
be assumed to be additive and independent of the basic motion.

\item[c)] If $\E[X_t^2]<\infty$, the covariance can be viewed as a limit of the codifference,
\bgeq
\lim_{\theta\to 0}\tau_X^\theta(t) = r_X(t),
\eeq
which stems from expanding the complex exponents in definition \Ref{eq:cdfDef}
into a Taylor series up to the second term and noting that we obtained the
logarithm of expression $(1+\theta^2 r_X(t)+o(\theta^2))^{\theta^{-2}}$. It is
then justified to treat the codifference as a generalisation of the covariance.

\item[d)] For a Gaussian process the codifference equals the covariance for any $\theta$
\bgeq
\tau_X^\theta(t)=r_X(t),
\eeq
which follows immediately from a short calculation, see
Eq. \Ref{eq:tauCexpl}. Therefore comparing the codifference and the covariance
can be used to measure non-Gaussianity.

\end{itemize}

One intuitive property, that the codifference does not have, is
symmetry. Considering two variables we fix the first one and negate the
second one ($x\mapsto -x$), and we expect the strength of dependence to be the
same but for the sign to change. This is the case for the covariance, but not
for the codifference, which is by design non-linear. Even in the borderline
case $X_{s+t}=-X_s$ we do not have a guarantee that $\tau_X^\theta(t)<0$,
counterexamples can be given even for the otherwise well-behaved class of
processes considered later. It is actually possible to remove this sometimes
inconvenient property by introducing the symmetrised codifference
\bgeq\label{eq:symmCodDef}
\widetilde\tau_X^\theta(t) \defeq \f{1}{2\theta^2} \ln\f{\E\lt[\e^{\I\theta(X_{s+t}-X_s)}\rt]}{\E\lt[\e^{\I\theta(X_{s+t}+X_s)}\rt]}
\eeq
which for all symmetric distributions changes sign with respect to
reflection, $X_{s+t}\mapsto -X_{s+t}$. This quantity can be useful if
one wants to compare the strength of positive and negative dependencies,
but there is a cost: the symmetrised codifference is "linear enough"
to ignore many types of non-linear ergodicity breaking, similarly to  the
covariance, see Eq. \Ref{eq:symmCod0}. For this reason further on we will
use the non-symmetrised codifference and study systems with a positive type
of dependence, at least in some suitable limit, such as $t\to\infty$.

Note that if the codifference is a generalisation of the covariance, one should reasonably expect that there exists a generalisation of the MSD defined in a similar spirit. Indeed, let us consider the formula
\bgeq\label{eq:LCFDef}
\zeta_X^\theta(t)\defeq -\f{2}{\theta^2}\ln\E\lt[\e^{\I\theta(X_t-\mu_X(t))}\rt].
\eeq
This quantity may seem trivial, because studying the distribution in
Fourier space is a classical method of basic probability theory. But, the
distinguishing part of this definition is that the result is treated primarily
as a function of time and it is conveniently transformed, so that it can be
interpreted as a measure of dispersion with the same unit as $X^2$. Up to a
rescaling it can be considered a cumulant generating function calculated at
imaginary argument, but such a quantity does not seem to have an established
name in the literature, so we will call it by the straightforward term "log
characteristic function", in short LCF. It is clear that in analogy to the
features of the codifference, the LCF measures mainly the spread of the bulk
of the probability and is much less influenced by the distribution's tails
than the MSD. As before, the first factor, here $2/\theta^2$, is optional
and only needed when one wants to compare the LCF to the MSD.

The LCF is indeed a reasonable measure of dispersion, as shown by the following properties:
\bg{itemize}
\item[a)] For independent $Y_t, Z_t$ and $X_t=Y_t+Z_t$,
\bgeq
\zeta_X^\theta(t) = \zeta_Y^\theta(t) + \zeta_Z^\theta(t).
\eeq
\item[b)] For any Gaussian process the LCF equals the MSD,
\bgeq
\zeta_X^\theta(t) =\delta^2_X(t).
\eeq
\item[c)] As we stretch the probability density of $X_t$, the LCF diverges, that is,
\bgeq
\lim_{c\to\infty}\zeta_{cX}^\theta(t) = -\f{2}{\theta^2} \ln \lim_{c\to\infty} \E\lt[\e^{\I\theta c X_t}\rt] = -\f{2}{\theta^2} \ln 0^+= \infty.
\eeq

\end{itemize}

The first two facts are analogues of the corresponding properties of the
codifference which allow one to trace the influence of the noise and detect
non-Gaussianity. The point c) is just the Riemann-Lebesgue lemma in disguise:
it corresponds to the intuition that the rescaled process should have a larger
spread. It should be mentioned that in general the LCF can be negative or
complex valued, which is highly undesirable. However, for the considered
models, which are based on internal Gaussian dynamics, this will never be
the case, as proved in Proposition \ref{prp:basic}.

Decomposing any process with independent increments into a sum of its jumps shows that in this case $\zeta_X^\theta(t)$ is a linear function. In particular, this holds of L\'evy flights \cite{taqqu}. It also holds for continuous time random walks with exponential waiting times \cite{Codling}, for which
\bgeq
\zeta_X^\theta(t) = \f{1-\EE{\e^{\I\theta J}}}{2\theta^2\E[T]} t,
\eeq
where $J$ is one jump and $T$ is one waiting time of diffusion $X$. The dependence on $T$ is the same as for the MSD,
\bgeq
\delta_X^2(t) = \f{\E[J]}{\E[T]} t,
\eeq
only the scaling depending on $J$'s distribution changes from non-linear to linear.

The LCF can also be used for finite- or infinite-variance models which are "anomalous" in some sense. A basic example is fractional L\' evy stable motion $L_\alpha^H$ \cite{selfSimilarStable}. It is stable and self-similar which implies that
\bgeq
\zeta_{L_\alpha^H}^\theta(t) = C_\theta t^{\alpha H},
\eeq
for some constant $C_\theta$, which depends on the chosen normalisation. This formula agrees with the intuition that a measure of the spread in this case should behave like a power law. Somewhat surprisingly, the situation is different for continuous time random walks with power-law waiting times, which are used to model subdiffusion. Such processes after rescaling converge to subordinated Brownian motion $B(S_\alpha(t))$, for which the LCF can be calculated directly, using the well-known properties of the inverse $\alpha$-stable subordinator $S_\alpha$ \cite{subord},
\bgeq
\zeta_{B(S_\alpha)}^\theta(t) = -\f{2}{\theta^2} \ln E_\alpha\lt(-\f{\theta^2}{2}t^\alpha\rt),
\eeq
where $E_\alpha$ is the Mittag-Leffler function \cite{haubold}. This function approaches infinity like a logarithm; the exact asymptotic is shown in Eq. \Ref{eq:gBM}. The difference between these two models of anomalous diffusion is that $L_\alpha^H$ is self-similar, so its PDF spreads in the uniform manner, whereas for $B(S_\alpha)$ the bulk is much more constrained than the tails.

After this brief discussion about the general properties of the codifference
and related notions, we will study its behaviour in more detail for models
based on random parameters of motion and for models based on random and
time-varying diffusion coefficient. The next section (B) provides a general
physical overview and concrete examples useful for the modelling. The third
and the last section (C) is dedicated to presenting mathematical results and
calculation techniques. The paper is written such that, if the reader prefers,
the physical and mathematical sections B and C can be read independently.

\setcounter{equation}{0}

\section{Modelling}

\subsection{Gaussian diffusion governed by random parameters}

One of the core concepts behind ergodicity and ergodicity breaking is the idea
of looking at information contained in a single trajectory. We speak about
ergodicity if the data that can possibly be gained analysing one, sufficiently
long, series of observations, is the same as if one analyses all possible
trajectories in the ensemble \cite{modErg}. Conversely, if this amount of
information is smaller, we speak about ergodicity breaking. In other words,
there is some information contained in a given trajectory, and using only a
single trajectory we
omit the amount contained in the rest. This is sometimes also rephrased as
confinement in the phase space, but this language must be used carefully as
the said space has a subtle  structure.\footnote{Even for classical Brownian
motion it is the infinitely dimensional Wiener space \cite{janson}}

From a different perspective, modelling based on the information content
often leads to an intuitive description, because the differences between
trajectories often stem from differences between diffusing particles
and differences between their local surroundings. Both may occur, e.g., in
biological systems. The latter case requires the additional assumption that
the inhomogeneity present in the surroundings varies on a length scale
of the mean distance between trajectories, but does not vary much at the
scale of the trajectories themselves. That is, distinct trajectories have
distinct surroundings, but each particle is sufficiently localised so that
the state of the medium around it does note change significantly. This is
reasonable for example when the particles are trapped or the measurement
time is sufficiently short---compare, e.g., the absolute spread of the traced
particles in \cite{weakErgJeon}.

In any case, this information can be parametrised, which leads to the so-called
hierarchical or multilevel modelling \cite{multilevel}, which in the context of
physics is also called "superstatistics" (a short term for "superposition
of statistics") \cite{beck2}. Deterministic parameters of the basic model
become random on an additional statistical layer.

\subsubsection{Random diffusion coefficient.}

For diffusion the simplest example of an hierarchical model is the motion with
a random diffusion coefficient, the situation when different trajectories
depict movements with varying average mobilities. A typical model of such
observations is the grey Brownian motion \cite{grey1,grey2,grey3}
\bgeq
B_{2H,\beta}(t) = \sqrt{D_\beta} B_{H}(t).
\eeq
Here $B_H$ is fractional Brownian motion \cite{mandelbrot} and the diffusion coefficient $D_\beta$ is an independent random variable with the so-called $\beta$ M-Wright distribution \cite{MWright}. The moments of grey Brownian motion are the same as those of fractional Brownian motion up to a multiplicative constant, therefore the MSD still grows as $t^{2H}$ and the process models anomalous diffusion. Nevertheless, a straightforward calculation yields that the LCF can be expressed using the Mittag-Leffler function,
\bgeq
\zeta_{B_{2H,\beta}}^\theta(t) = -\f{2}{\theta^2}\ln E_\beta\lt(-\f{\theta^2}{2}t^{2H}\rt) \sim \f{1}{\Gamma(\beta+1)}t^{2H},\quad t\to 0^+,
\eeq
which also yields
\bgeq\label{eq:gBM}
\zeta_{B_{2H,\beta}}^\theta(t)=\f{4H}{\theta^2} \ln t + \f{2}{\theta^2}\ln\lt(\f{\theta^2\Gamma(1-\beta)}{2}\rt) + o(1),\quad \beta\neq 1, t\to\infty.
\eeq

Here the asymptotic `$+ o(1)$' is pointwise, which is stronger than the asymptotic proportionality `$\sim$'; in the sense of `$\sim$' the term $4H/\theta^2 \ln t$ is dominating and the logarithmic behaviour clearly distinguishes the LCF from the power-law MSD at long times. This crossover behaviour can be used to distinguish grey Brownian motion from fractional Brownian motion (case $\beta = 1$ \cite{grey1}) and diffusing-diffusivity model (Eqs. \Ref{eq:zetaLim} and \Ref{eq:zetaLim2}). The very slow log increase of the LCF is not surprising: because the diffusion constant is random, but fixed and it constrains the relaxation of the probability density---it is detected by the LCF, but ignored by the MSD; for a more general result see Proposition \ref{prp:randDLCF} d).

Grey Brownian motion models free, unconfined movements and is therefore not stationary. Still, the codifference can be used for its  increments $\Delta B_{2H,\beta}(t)\defeq B_{2H,\beta}(t+\Delta t)-B_{2H,\beta}(t)$. The calculation is again not hard and yields
\bgeq
\tau_{\Delta B_{2H,\beta}}^\theta(t)=\f{1}{\theta^2}\ln\f{E_\beta\lt(-\theta^2\lt(\Delta t^{2H}-(|t+\Delta t|^{2H}+|t-\Delta t|^{2H})/2\rt)\rt)}{E_\beta\lt(-\theta^2\Delta t^{2H}/2\rt)}.
\eeq
The covariance decays to zero like a power law $t^{2H-1}$, but the function above decays to the non-zero constant
\bgeq
\tau_{\Delta B_{2H,\beta}}^\theta(\infty)=\f{1}{\theta^2}\ln\f{E_\beta\lt(-\theta^2\Delta t^{2H}\rt)}{E_\beta\lt(-\theta^2\Delta t^{2H}/2\rt)}.
\eeq
 This means that there is some degree of dependence left even at $t=\infty$ which the covariance does not detect, but the codifference does. Indeed, it can be interpreted as a joint dependency on the trajectory-wise fixed but random diffusion coefficient $D_\beta$.

The above simple example shows that the codifference does not directly detect non-ergodicity, it rather detects dependence. The notion of mixing is useful to describe this idea. It is a property which states that the future evolution of the process after a long delay becomes independent of its past values. Formally speaking, the process is mixing when, if we calculate some statistic in some finite time interval starting at $s$, and later on any other statistic starting at $s+t$, these two must become independent as $t\to\infty$ \cite{ergTh}. Therefore, analysing the codifference, which measures the dependence between $\exp(-\I\theta X_s)$ and $\exp(\I\theta X_{s+t})$, allows one to exclude mixing, i.e. to indicate the presence of a non-vanishing dependence. The latter means that the motion is constrained in phase space, which in turn implies ergodicity breaking.\footnote{The remaining class of processes which are ergodic but non-mixing is complicated and those do not seem to appear in applications. For a mathematically constructed example of such a process and the discussion see \cite{gaussErg}.}

Thus, for a very large class of systems one does not need to study time-averages to detect non-ergodicity. It is sufficient to find a proper memory function which will indicate non-mixing. As we demonstrate the covariance fails in this role for the considered models, but the codifference works.

These detecting capabilities of the codifference work under quite general circumstances. If we observe any ensemble of mixing, zero mean Gaussian trajectories, the covariance will converge to zero. This happens because for Gaussian process, mixing is equivalent to a decay of the covariance \cite{ergTh,gaussErg}, and the mixture of decaying covariance functions is decaying. But, the ensemble of trajectories as a whole will not be ergodic, which will not be detected by the covariance. Let $\mathcal C$ is some parametrisation of this mixture, then the conditional average $D = \E[X_t^2|\mathcal C]$ be the resulting, possibly random, conditional variance. We call it $D$ because if the data $X$ corresponds to the velocity or increments of displacements, it will be proportional to the diffusion coefficient. Under these assumptions the codifference converges to the constant
\bgeq\label{eq:codCLim}
\tau_X^\theta(\infty)=\f{1}{\theta^2}\ln\f{\E\lt[\e^{-\theta^2 D}\rt]}{\E\lt[\e^{-\theta^2 D/2}\rt]^2}\ge 0,
\eeq 
as proven in Proposition \ref{prp:codInf}. This quantity is related to the coefficient of variation defined as the standard deviation divided by the mean \cite{statDict}. Denoting it by $\mathrm{CV}[X]$, the formula above can be expressed as $\theta^{-2}\ln(\mathrm{CV}[\exp(-\theta^2D/2)]^2+1)$ which is an increasing function of $\mathrm{CV}[\exp(-\theta^2D/2)]$ and asymptotically quadratic for small $\mathrm{CV}$. The coefficient of variation is a measure of dispersion, hence so is $\tau_X^\theta(\infty)$ which reflects the randomness of $D$. This behaviour is also equivalent to detecting a residual dependence and the resulting non-mixing/non-ergodicity.

Outside of the useful limit $t=\infty$ not much can be said about the properties of the codifference in such a wide and general class. The situation changes if we consider a more specific model. The idea behind grey Brownian motion and many works about superstatistics \cite{beck2} is that the trajectories differ mainly by the diffusion coefficient, other properties are not significantly distinct. A simple model of such a system can be written as
\bgeq
X_t = \sqrt{D}Y_t.
\eeq
We assume that the process $Y$ describes the joint form of dependence common for all trajectories. We consider a Gaussian $Y$, which for grey Brownian motion would be fractional Brownian motion. Another reasonable choice would be, e.g., a solution of the Langevin equation. In this case, as long as $Y$ is stationary (i.e., for free diffusion we consider increments or the velocity process), the covariance is
\bgeq
r_X(t)=\E[D]r_Y(t),
\eeq
of course as long as $\E[D]<\infty$. If the process $Y$ has sufficiently long memory, $r_Y(t)\approx 0$ in the considered time scale, also $r_X(t)\approx 0$. The covariance does not detect the additional dependence introduced by random $D$.

At the same time the codifference can be expressed as a function of the covariance of $Y$, precisely as
\bgeq
\tau_X^\theta(t) = \f{1}{\theta^2}\ln\f{\E\lt[\e^{-\theta^2D(1-r_Y(t))}\rt]}{\E\lt[\e^{-\theta^2D/2}\rt]^2}
\eeq
for any $D$, no matter if $\E[D]<\infty$. It clearly converges to the constant \Ref{eq:codCLim} as $r_Y(t)\to 0$ and detects the additional non-linear dependence.

For a general, possibly non-stationary $Y$ with $\E[Y_t^2]=\delta_Y^2(t)$, the representation of the LCF is
\bgeq
\zeta_X^\theta(t) = -\f{2}{\theta^2}\ln\E\lt[\e^{-\theta^2D\delta_Y^2(t)/2}\rt].
\eeq
Given some model of $D$ these formulae can be made completely explicit, examples are given in Table \ref{tab:cdf}. The first example is the gamma distribution $D\deq \mathcal G(\alpha,\beta)$ in which the coefficient $\alpha$ describes the power-law behaviour of the PDF near 0 and $\beta$ is the rate of exponential decay of the tails (the specific case $\mathcal G(1,\beta)$ is the exponential distribution); it models common types of experiments in which the distribution of diffusion coefficients resembles a bump concentrated around some finite constant and high values of $D$ become exponentially less probable. This case is also illustrated in Figure \ref{fig:codCmp}.

\bg{figure}
\centering
\includegraphics[width=12cm]{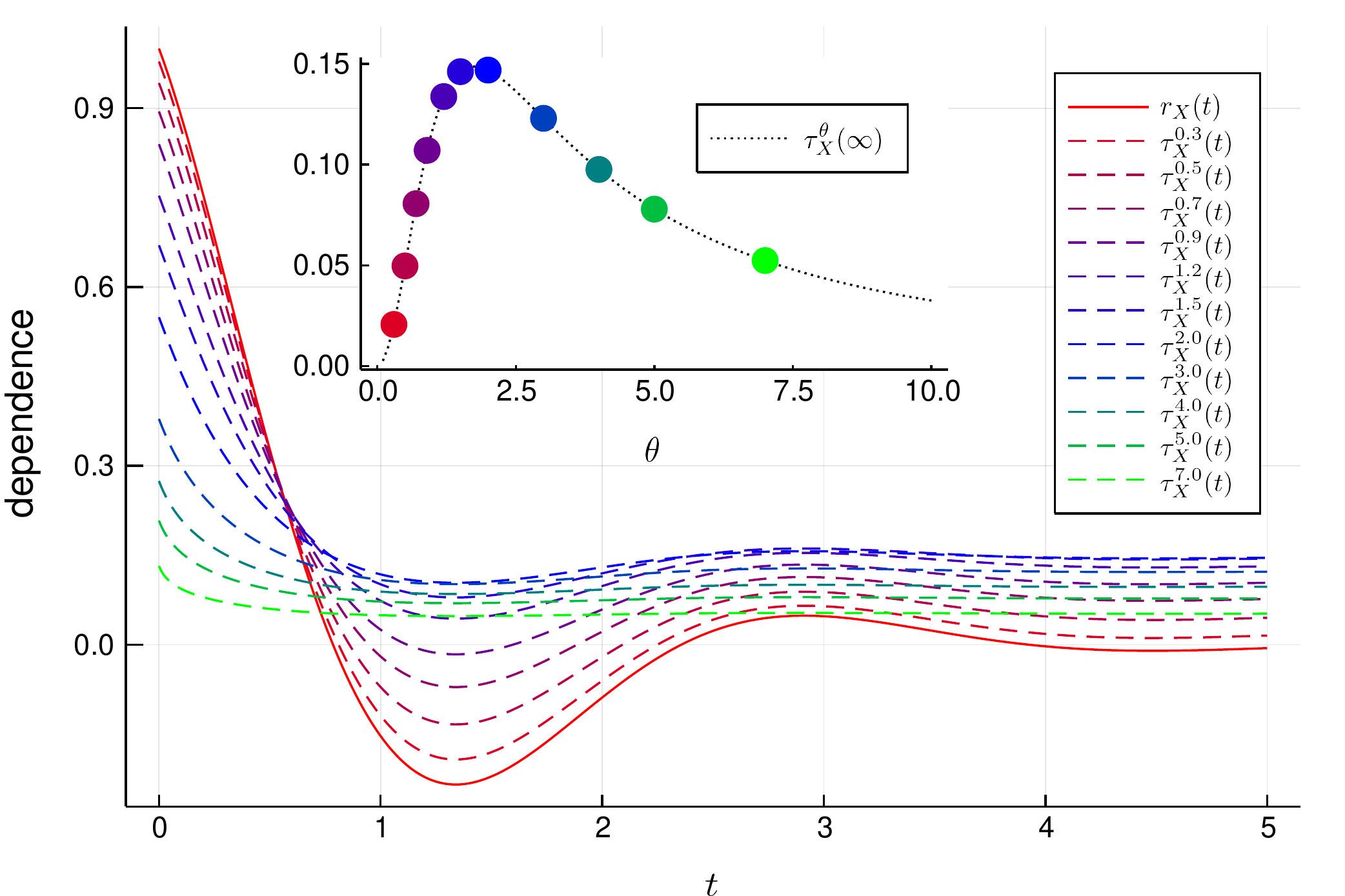}
\caption{Codifference $\tau$ and covariance $r$ of the process $X_t=\sqrt{D}Y_t$ with $D\deq \mathcal G(1,1)$ and $r_Y(t)=\cos(t)\exp(-t)$, as given by Table \ref{tab:cdf}. Various properties of the codifference are visible: for $\theta\to 0$ it converges to the covariance; the codifference and the covariance increase and decay in the same intervals; at $t=0$ the codifference is smaller than the covariance; as $t\to \infty$ the codifference converges to a $\theta$-dependent value $\tau_X^\theta(\infty)$ which is a functional of the law of $D$; the type of asymptotic of $r_X(t)$ and $\tau_X^\theta(t)-\tau_X^\theta(\infty)$ is the same (here: exponential decay). The derivations are presented in Proposition \ref{prp:randD}. }\label{fig:codCmp}
\end{figure}

Diffusion coefficients with a heavy-tailed distribution result in a motion that itself exhibits heavy tails of the PDF, a phenomenon actively investigated in transport, finance, turbulence and many other systems \cite{restaurant,klafterBeyond,light}. A classical model of this case is  the one-sided $\alpha$-stable subordinator $\mathcal S(\alpha,c)$, determined by its Laplace transform $\exp(-(cs)^\alpha)$. The resulting type of process was thoroughly studied in the literature concerned with stable distributions \cite{taqqu}. This process is called sub-Gaussian, which is arguably a confusing term. In this case the process $X$ has no second moment, therefore attempts to estimate its covariance will lead to a diverging result. This is visible in the formulae for the codifference and the LCF, which diverge as $\theta\to 0$. But, for any $\theta >0$ the codifference and the LCF are finite and can be estimated in a standard way, and from the result if one wishes the covariance and the MSD of $Y$ can be reconstructed.

For a distribution concentrated around its mean value one can use Gaussian $\mathcal N(\mu,\sigma^2)$ or uniform $\mathcal U(a,b)$ distributions, however the former is only a valid model for $\sigma \ll \mu$, when the probability that $D<0$ can be neglected. 

\begin{table}
	
	\begin{tabular}{ r || c | c }
			law of $D$	 & codifference $\tau_X^\theta(t)$ & LCF $\zeta_X^\theta(t)$ \\ \hline\hline
		$\mathcal G(\alpha,\beta)$  & $\f{\alpha}{\theta^2}\ln\f{\lt(1+\theta^2/(2\beta)\rt)^2}{1+\theta^2(1-r_Y(t))/\beta}$ & $\f{2\alpha}{\theta^2}\ln\lt(\f{\theta^2}{2\beta}\delta_Y^2(t)+1\rt)$ \\ \hline	
		
		$\mathcal S(\alpha,c)$  & $c^\alpha\theta^{2\alpha-2}\lt(2^{1-\alpha}-\lt(1-r_Y(t)\rt)^\alpha\rt)$ & $2^{1-\alpha}c^\alpha\theta^{2\alpha-2} \lt(\delta_Y^2(t)\rt)^\alpha$\\  \hline
		
		$\mathcal N(\mu,\sigma^2)$  &$\mu r_Y(t)+\f{(\theta\sigma)^2}{2}(1-r_Y(t))^2-\mu-\lt(\f{\theta^3}{8}\sigma^2-\f{\theta}{2}\mu\rt)^2$ & $\mu\delta^2_Y(t)-\f{(\theta\sigma)^2}{4}(\delta_Y^2(t))^2$\\ \hline
		$\mathcal U(a,b)$  &$a r_Y(t)+\f{1}{\theta^2}\ln\lt(\f{\theta^2(b-a)}{4(1-r_Y(t))}\f{1-\e^{-\theta^2(b-a)(1-r_Y(t))}}{\lt(1-\e^{-\theta^2(b-a)/2}\rt)^2}\rt)$ & $a\delta^2_Y(t) -\f{2}{\theta^2}\ln\lt(\f{2\lt(1-\e^{-\theta^2(b-a)\delta^2_Y(t)/2}\rt)}{\theta^2\delta^2_Y(t)(b-a)}\rt)$\\
	\end{tabular}
		\caption{Formulae for the codifference and the LCF corresponding to common models of $D$: gamma, one-sided stable, Gaussian and uniform.}\label{tab:cdf}
\end{table}		

Even if the precise model of $D$ is not known, quite a lot can be said about the behaviour of the codifference. In Proposition \ref{prp:randD} we show that
\bg{itemize}
\item[a)] The codifference is a monotonic function of the covariance. If one increases, the second one also increases, the same goes for decreases.
\item[b)] If $\E[D]<\infty$ the codifference is smaller than the covariance for strong positive correlation, but larger for weak or negative correlations.
\item[c)] The approach to the value $\tau_X^\theta(\infty)$ has the same asymptotic as the decay of the covariance
\bgeq
\tau_X^\theta(t)-\tau_X^\theta(\infty)\sim \f{\E\lt[D\e^{-\theta^2 D}\rt]}{\E\lt[\e^{-\theta^2 D}\rt]}r_Y(t),\quad t\to\infty,
\eeq
assuming $r_Y(t)\to 0$, which is a typical case.
\end{itemize}

These are all desirable properties: the memory structure of the internal process $Y$ is reflected in a straightforward manner by the codifference. For small values of the covariance their relation is even linear, as stated in c), and the proportionality constant is finite for any distribution of $D$, due to the  truncating factor $\exp(-\theta^2 D)$.
 
Another property is that the codifference depends additively on $D$. Precisely speaking, if we decompose $D=D'+D''$ for some independent $D'$ and $D''$, the codifference also decomposes for
\bgeq
\tau_X^\theta(t) = \tau_{X'}^\theta(t) + \tau_{X''}^\theta(t),
\eeq
where $X'$ and $X''$ are processes with diffusion coefficients $D'$ and $D'$ respectively. Therefore subtracting the codifferences estimated from different samples may be used to analyse different sources of diffusivity. The derivation is given in Proposition \ref{prp:randD}.

Analogous features can also be checked for the LCF (Proposition \ref{prp:randDLCF}), which can also be decomposed for $D=D'+D''$ and is a monotonic function of the MSD, but is always smaller than the MSD, therefore detecting the additional constraints of the motion introduced by a random $D$.

At the end of the discussion about random diffusion coefficients we note that the behaviour of the codifference near $t=0$ can also give valuable information. In Proposition \ref{prp:codShortTime} we prove that for a typical case when $\E[D]<\infty$ its asymptotic reflects that of the covariance. However, if $\E[D]=\infty$ and $D$ has power tails, corresponding to the presence of high-volatility trajectories, the asymptotic of the codifference has an additional power law. As for Gaussian processes the behaviour of the covariance near $t=0$ is determined by their fractal dimension \cite[chapter 8.8]{adler}, the same is true for the codifference, which can be applied also for processes with no moments.

\subsubsection{Random memory decay rate.}

Another interesting type of models are ensembles of particles for which the time dependence may vary from trajectory to trajectory. The simplest model of a time-varying dependency is the exponential decay $ \exp(-t\Lambda)$, which is the covariance of Ornstein-Uhlenbeck process \cite{OUoryg}. It models many kinds of linear relaxation disturbed by additive noise. It was also studied as a model of the additive measurement noise itself \cite{Bryson,JeonNoise}. In the hierarchical model the decay rate $\Lambda$ may be random. The covariance of the resulting mixture of Ornstein-Uhlenbeck type trajectories was studied in \cite{superstatLang} in the context of a randomly parametrised Langevin equation.

The coefficient $\Lambda$ has a different physical interpretation depending on the details of the studied phenomenon. For the velocity of a Brownian particle it is proportional to the friction coefficient and its randomness is related to local changes of the viscosity and/or different shapes of the diffusing particles \cite{zwanzig}; in this system the fluctuation-dissipation relation also links the scaling to the temperature. For trapped particles $\Lambda$ is proportional to the stiffness of the confining harmonic potential (the prominent example being optical tweezers \cite{norregaard,ashkin}), therefore the randomness of $\Lambda$ is equivalent to an ensemble of traps with varying sizes, which are proportional to $\Lambda^{-1}$. 

Another case worth mentioning is that of viscoelastic anomalous diffusion \cite{goychuk}, for which the velocity (or increments) have power-law dependence $\propto t^{2H-1}$. This function can be expressed as $\exp(-\ln(t)(1-2H))$. Therefore it is enough to replace $t$ with $\ln t$ and the results further on will also follow for the ensemble of power-law memory trajectories characterised by random parameter $(1-2H)$. It is worth to note that the variability of the of the Hurst index $H$ seems to be more of a rule than an exception for biological systems \cite{nature2017,thapa,thapa1}.

We do not want to make the discussion overly technical, so below we will analyse only the case of deterministic scaling and random decay rate, $r_X(t|\mathcal C)=\sigma^2\exp(-t\Lambda )$. Results for more general $Df(\Lambda)\exp(-t\Lambda )$ are presented in Propositions \ref{prp:randLang}, \ref{prp:increments} and \ref{prp:generalRandDL}, which prove that the randomness of the scaling is not essential for most of the properties discussed below. We also note that sometimes one can remove the random scaling and normalise the trajectories using the estimate of scaling obtained from the Birkhoff ergodic theorem \cite{ergTh},
\bgeq\label{eq:estD}
r_X(0|\mathcal C)=\lim_{T\to\infty}\f{1}{T}\int_0^T\dd t\ X(t)^2.
\eeq
However, this procedure requires having access to sufficiently long trajectories.

A particular property of ensembles with fixed scaling is that any marginal distribution is Gaussian, i.e., all variables $X_t$ have Gaussian distribution with variance $\sigma^2$. But the codifference can be found to be
\bgeq\label{eq:cdfSimpl}
\tau_X^\theta(t)=\f{1}{\theta^2}\ln\EE{\e^{(\theta\sigma)^2\e^{-t\Lambda}}},
\eeq
and because it does not equal the covariance, the process as a whole is not Gaussian. The codifference indicates the presence of subtle non-Gaussianity of the memory structure. This formula can also be used to derive useful bounds between the codifference and the covariance, see Proposition \ref{prp:fixVar}.
 
Expanding in a Taylor series the exponent from \Ref{eq:cdfSimpl} leads to
\bgeq
\tau_X^\theta(t)=\f{1}{\theta^2}\ln\lt(1+\sum_{k=1}^\infty \f{(\theta\sigma)^{2k}}{k!}\E\lt[\e^{-kt\Lambda}\rt]\rt)\sim \f{1}{\theta^2}\sum_{k=1}^\infty \f{(\theta\sigma)^{2k}}{k!}\E\lt[\e^{-kt\Lambda}\rt],\quad t\to\infty.
\eeq
Note that $\sigma^2\E[\e^{-k t\Lambda}]=r_X(kt)$, so the result is a type of average over the values $r_X(kt)$. When the distribution of $\Lambda$ is not sufficiently concentrated near $0$ and the covariance decays fast (strictly speaking is rapidly varying \cite{deHaan,mikosch}), the term $k=1$ dominates the $t\to\infty$ asymptotic. This is the case, e.g., for the one-sided stable variable $\Lambda\deq\mathcal S(\alpha,c)$ for which
\bgeq
\tau_X^\theta(t) = \f{1}{\theta^2}\ln\lt(1+\sum_{k=1}^\infty\f{(\theta\sigma)^{2k}}{k!}\e^{-(c k t)^\alpha}\rt)\sim\sigma^2 \e^{-(c t)^\alpha},\quad t\to\infty
\eeq
that is, we observe a stretched exponential type of dependence.

When $\Lambda$ is more concentrated around 0 the situation differs. A basic example would again be the gamma distribution $\Lambda\deq\mathcal G(\alpha,\beta)$, for which
\bgeq
\tau_X^\theta(t)=\f{1}{\theta^2}\ln\lt(1+\sum_{k=1}^\infty\f{(\theta\sigma)^{2k}}{k!}\f{1}{(1+kt/\beta)^\alpha}\rt).
\eeq
When $\alpha=1$ (i.e., $\Lambda$ has an exponential distribution) the above can also be written using the incomplete gamma function. For any $\alpha$ all terms in the sum decay like $t^{-\alpha}$ and they are comparable. Because of this, the codifference also decays with the same power law, but the proportionality constant is non-trivial ,
\bgeq
\tau_X^\theta(t) \sim  \f{1}{\theta^2}\sum_{k=1}^\infty\f{(\theta\sigma)^{2k}}{k!}\f{1}{\lt(\f{1}{t}+k/\beta\rt)^\alpha}t^{-\alpha}\sim  \f{1}{\theta^2\beta^\alpha}\sum_{k=1}^\infty\f{(\theta\sigma)^{2k}}{k!}\f{1}{k^\alpha}t^{-\alpha},\quad t\to\infty.
\eeq

It is not surprising that this behaviour is not specific to a gamma distribution and can be observed for any $\Lambda$ with power-law PDF near $0^+$, see Proposition \ref{prp:randLang}. Similarly, if the PDF of $\Lambda$ decays fast near $0^+$, the codifference also decays fast. All these properties are analogous to those of the covariance \cite{superstatLang}, so here they can be used interchangeably or simultaneously, as a mean to obtain stronger statistical verification.

They are also similar in that both do not detect the non-ergodicity, more precisely the non-mixing, of this system. As was already demonstrated for the covariance it is a common occurrence resulting from its linearity. The codifference fails, because it does measure only a reduced form of mixing. For the process to be mixing it means that any two sets of multiple disjoint measurements must become asymptotically independent, i.e., the vectors $[X_{s_1},X_{s_2},\ldots,X_{s_n}]$ and $[X_{s_1+t},X_{s_2+t},\ldots,X_{s_n+t}]$ have to become independent as $t\to\infty$. The codifference (and for that matter also the covariance) measures only the dependence between two values $X_s$ and $X_{s+t}$.

For a process with a random decay rate these are asymptotically independent and the one-point distributions are relaxing. Therefore, in order to detect non-ergodicity, we need to analyse the dependence between at least three values. A practical choice is to use four values divided into two pairs $[X_s,X_{s+\Delta t}]$ and $[X_{s+t},X_{s+\Delta t+t}]$. The values in the first pair are correlated as $\e^{-\Delta t\Lambda}$ trajectory-wise, analogously for the values of the second pair. This property of both pairs is fixed and random, i.e., it is a constant of motion which can be detected. Probably the simplest method to achieve this is to calculate increments
\bgeq
\Delta X_t\defeq X_{t+\Delta t}-X_t
\eeq
and study the codifference of those. A short calculation given in Proposition \ref{prp:increments} shows that this method indeed works and
\bgeq\label{eq:dXcod}
\tau_{\Delta X}^\theta(\infty) = \f{1}{\theta^2}\ln\f{\EE{\e^{-2(\theta\sigma)^2\lt(1-\e^{-\Delta t \Lambda}\rt)}}}{\EE{\e^{-(\theta\sigma)^2\lt(1-\e^{-\Delta t \Lambda}\rt)}}^2} \ge 0.
\eeq
The result depends on $\Lambda$ in a complex manner, but it can be easily estimated numerically. We can also use the fact that for small $\Delta t$ the conditional covariance of increments is 
\bgeq
r_{\Delta X}(t|\Lambda,D) \in 2\Delta t^2 \sigma^2\Lambda^2\e^{-t\Lambda}+\mathcal O(\Delta t^4)
\eeq 
and normalise the process, $\Delta\widetilde X_t \defeq \Delta X_t/\sqrt{\Delta t}$. The result then simplifies and becomes independent of $\Delta t$,
\bgeq
\lim_{\Delta t\to 0^+}\tau_{\Delta\widetilde X}^\theta(\infty) = \f{1}{\theta^2}\ln\f{\EE{\e^{-2(\theta\sigma)^2\Lambda}}}{\EE{\e^{-(\theta\sigma)^2\Lambda}}^2}.
\eeq
We stress here that this method cannot be applied using the covariance, which, calculated from increments, decays to $0$ and does not detect this specific memory structure. Its decay is even quicker than for the original process and proportional to the power law decay $t^{-\alpha-2}$ \cite{superstatLang}. Intuitively speaking, the decay rate is quicker by a factor $t^{-2}$, because the scale of $\Delta X$ depends on $\Lambda$ as $\Lambda^2$ and the trajectories with stronger correlation have smaller amplitude and add less to the average. This property has its analogy for the codifference, for which $\tau_{\Delta X}^\theta(t)-\tau_{\Delta X}^\theta(\infty)$ also decays like $t^{-\alpha-2}$ (see Proposition \ref{prp:generalRandDL} for a more general result). This time the faster decay rate actually helps in detecting ergodicity breaking, making the limit $\tau_{\Delta X}^\theta(\infty)$ visible even at short times.  The numerical illustration of the discussed behaviour is shown in Figure \ref{fig:MC}.

\bg{figure}
\centering
\includegraphics[width=11cm]{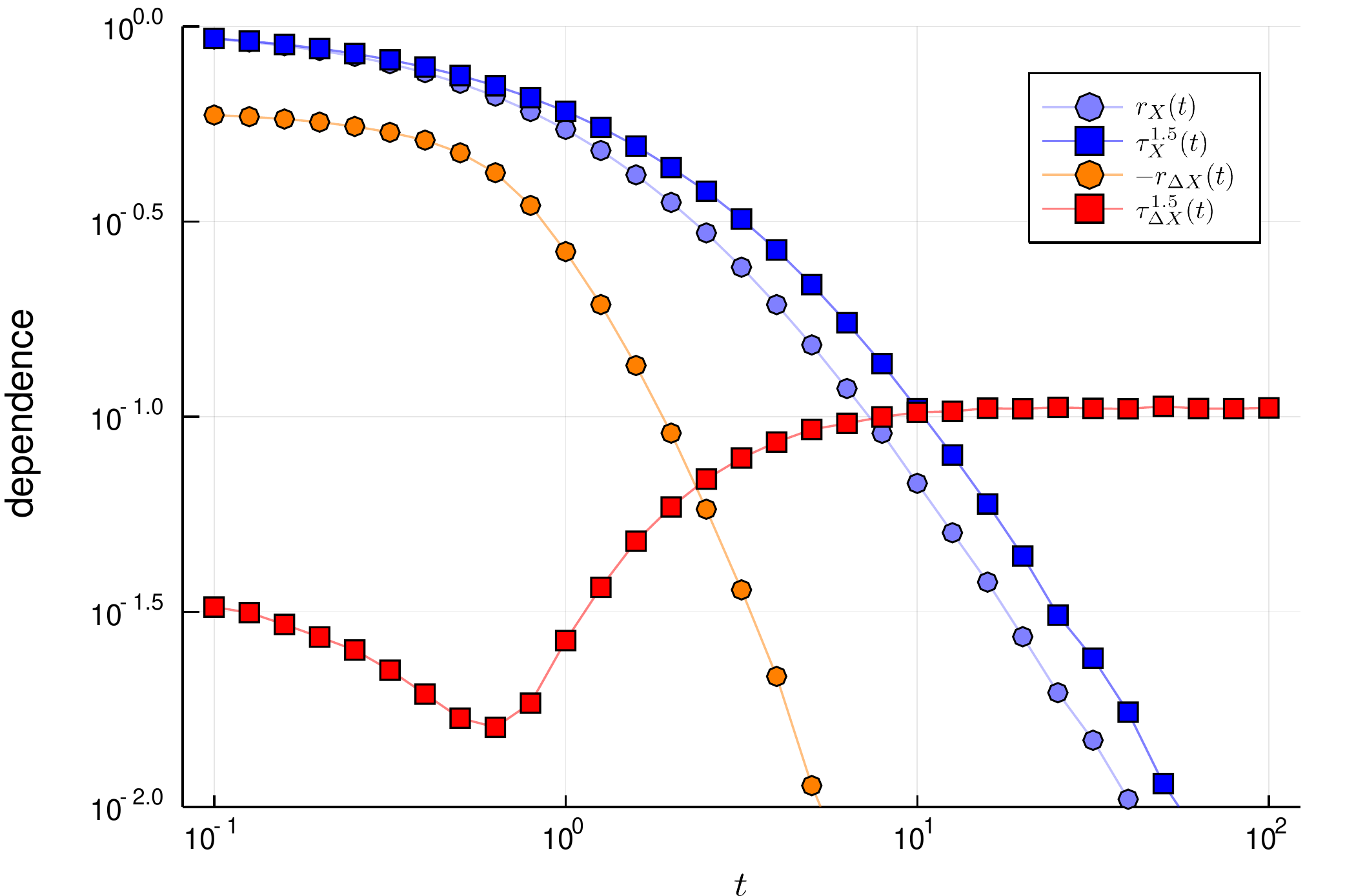}
\caption{Estimated codifference $\tau$ and covariance $r$
estimated from the process with random decay rate $\Lambda \deq \mathcal
G(3/2,1/2)$ and $\Delta t=1$. In the presented domain the covariance
$r_{\Delta X}$ was negative, so we plotted the negated value. One can
observe the predicted power law decays $\propto t^{-3/2}$ and $\propto
t^{-3/2-2}$ (Eq. \Ref{eq:langTauAs}); the codifference of increments
detects the non-ergodicity by converging to a constant $\tau_{\Delta
X}^{1.5}(\infty)\approx 0.105$, which fits perfectly Eq. \Ref{eq:dXcod}. The
value $\theta=1.5$ was chosen to best illustrate the interesting properties;
for smaller $\theta$ the codifference $\tau_X^\theta$ becomes closer to
the covariance $r_X$, for larger $\theta$ the codifference $\tau_{\Delta
X}^\theta$ converges faster to the $t\to\infty$ limit. To
present smooth curves in the whole presented range we used a large $10^7$
sample; the general shape of the presented functions is already visible for samples
around $10^4$; a significant difference between $r_{\Delta X}$ and $\tau_{\Delta
X}$ is observed using even a few hundred trajectories. Examples for smaller
sample sizes are presented in figure \ref{addfig}.}
\label{fig:MC}
\end{figure}

\subsection{Diffusing-diffusivity}

In the preceding sections we considered models which were non-Gaussian and non-ergodic. For non-Gaussian but ergodic models the codifference can also be a useful measure of dependence. In particular we show that it can be successfully used to analyse diffusing-diffusivity models. We now assume that the increments of $X_t$ are Brownian fluctuations, but rescaled by a time-dependent random diffusivity $D_t$,
\bgeq \label{eq:dXDef}
\dd X_t = \sqrt{D_t}\dd B_t.
\eeq
This is a generalisation of the random parameter model, for which $D_t=\const$ Because we modified the dynamical equation by replacing the previously constant parameter with a stochastic process, models of this class are sometimes called "doubly stochastic" \cite{doubleStoch}. Before application in  physics, they were extensively used in financial engineering, where it is natural to assume that parameters of the market, such as the volatility, vary in time. In 1985 Cox, Ingersoll and Ross \cite{CIR} proposed a model  of interest rate (now commonly named CIR), which describes a non-negative stochastic process with linear mean-reverting property. In 2012 Chubynsky and Slater  independently proposed a special case of the CIR process as a model of non-Gaussian diffusion \cite{diffDiff, Jain2017}. This led the way to a wider range of models based on fluctuating diffusivity coefficient with a short time memory \cite{Jain16,Tyagi, Chechkin2017, Sposini18,denis}. The evolution of the diffusion coefficient in the CIR model is defined by the stochastic equation
\bgeq\label{eq:DDef}
\dd D_t = a(b - D_t)\dd t + \sigma \sqrt{D_t}\dd B_t,
\eeq
where $a>0$ describes the speed of return to the mean $b>0$, and $\sigma>0$ regulates the amplitude of the fluctuations. In this equation as $D_t\to 0$ the term $a(b-D_t)\dd t\approx ab\dd t>0$ starts to dominate the fluctuations with the mean-squared amplitude $\E[(\sqrt{D_t}\dd B_t)^2]= D_t\dd t$, consequently $\dd D_t>0$ which causes the motion to stay positive. We assume that the system evolved for a long time before the start of the measurement and has reached the stationary gamma distribution $D_0\deq\mathcal G(2ab/\sigma^2,2a/\sigma^2)$ \cite{finMeth}.  Because of the non-Gaussianity the LCF function should differ from the MSD. Conditioning by $D_t$, it can be expressed by the formula
\bgeq\label{eq:diffDiffZeta}
\zeta_X^\theta(t)=-\f{2}{\theta^2}\ln\EE{\exp\lt(-\f{\theta^2}{2}\int_0^t\dd s\ D_s\rt)}.
\eeq
Expanding the above  in powers of $\theta^2$ shows that again $\zeta_X^\theta(t)\to \delta_X^2(t)$ as $\theta\to 0$.

The average in \Ref{eq:diffDiffZeta} appears in the calculation of the expected price of zero-coupon bond and was calculated in the initial paper of Cox, Ingersoll and Ross \cite{CIR}, who derived the differential equation which it fulfils and then solved it; a more general result is also available in \cite{finMeth}. The calculation was performed for the case when $D_0$ is fixed and deterministic, however their result can be easily extended for stationary $D$ by averaging over the equilibrium $\mathcal G(2ab/\sigma^2,2a/\sigma^2)$ distribution of $D_0$. Then the formula for the LCF reads
\bgeq\label{eq:zetaDD}
\zeta^\theta_X(t) = \f{4 ab}{(\theta\sigma)^2}\ln\lt(  \lt(\f{1}{2}+\f{(\theta\sigma)^2}{4\gamma_\theta a} + \f{a}{2\gamma_\theta}\rt)\e^{(\gamma_\theta-a)t/2} + \lt(\f{1}{2}- \f{(\theta\sigma)^2}{4\gamma_\theta a}  -\f{a}{2\gamma_\theta}\rt)\e^{-(\gamma_\theta+a)t/2}      \rt)
\eeq
with $\gamma_\theta=\sqrt{a^2+(\theta\sigma)^2}$. From that a brief calculation proves that the motion is Fickian for long times
\bgeq\label{eq:zetaLim}
\zeta^\theta_X(t) \in \f{2 ab}{(\theta\sigma)^2}(\gamma_\theta-a) t + \f{4 ab}{(\theta\sigma)^2}\ln \lt(\f{1}{2}+\f{(\theta\sigma)^2}{4\gamma_\theta a} + \f{a}{2\gamma_\theta}\rt) + o(1), \quad t\to\infty,
\eeq
and also for short time, albeit with a diffusion scale agreeing with the MSD
\bgeq \label{eq:zetaLim2}
\zeta_X^\theta(t)\sim b t,\quad t\to 0^+ ,
\eeq
which should come as no surprise. For an illustration of these formulae see Figure \ref{fig:LCF}, where we present results of Monte Carlo simulations compared to the theoretical predictions. See also the crossover behaviour of the MSD in the random diffusivity model in \cite{Sposini18}.

\bg{figure}
\centering
\includegraphics[width=12cm]{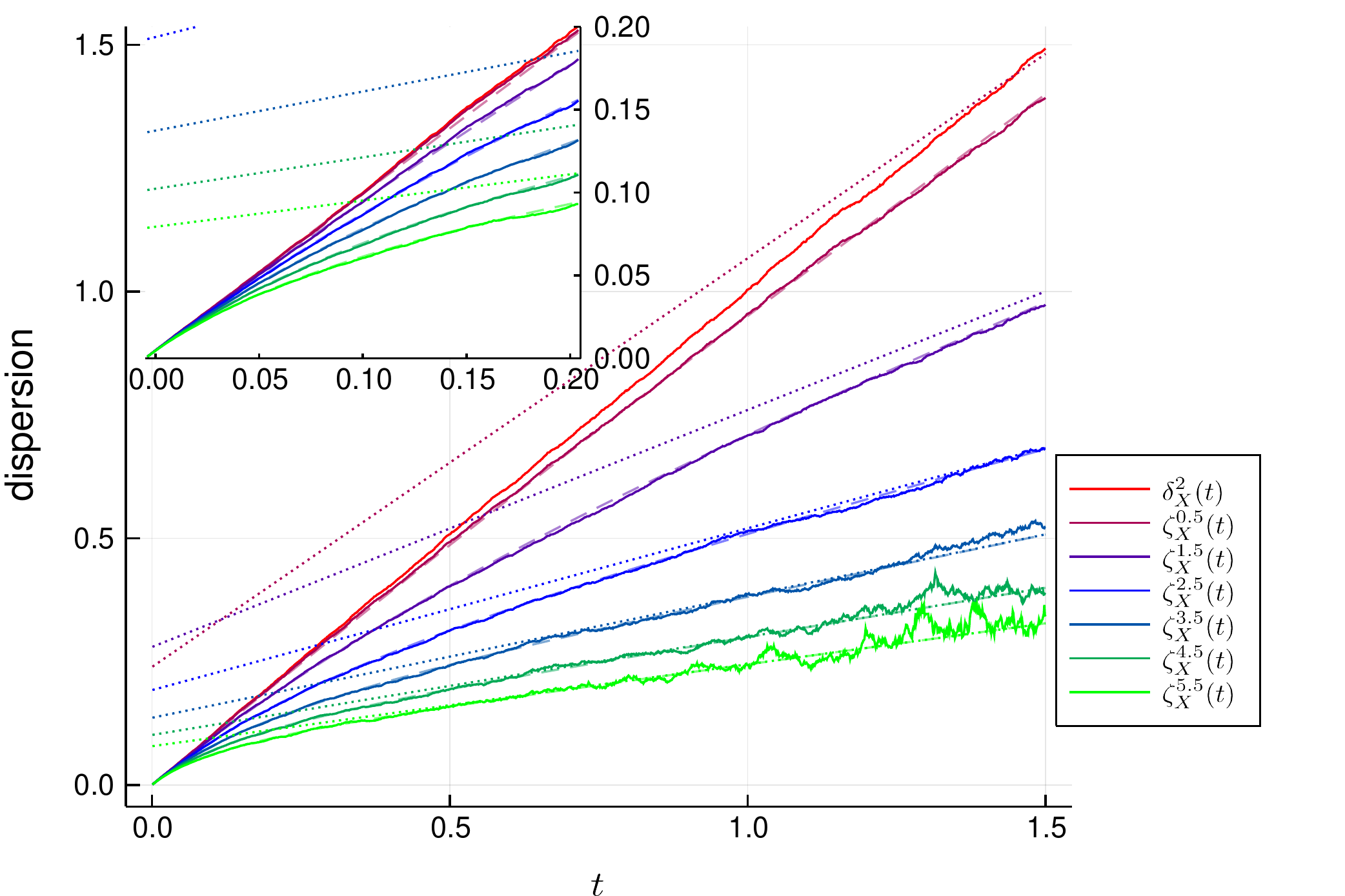}
\caption{MSD and LCF for the diffusing-diffusivity CIR model defined by \Ref{eq:dXDef} and \Ref{eq:DDef} with $a=1/2,b=1,\sigma=1$. Solid lines are functions estimated from $2\times10^4$ trajectories simulated using the Euler scheme with $\Delta t = 10^{-3}$; dashed lines are the analytical predictions given by \Ref{eq:zetaDD}; dotted lines are the long-time linear limits \Ref{eq:zetaLim}. It is clearly observed that the MSD exhibits a single linear law whereas the LCF switches between two linear laws at $t\to 0^+$ and $t\to\infty$. Also note that for large $\theta$ and $t$ the estimation becomes unstable. It is caused by $\E[\cos(\theta X_t)]$ becoming comparable in amplitude to the estimation uncertainty; for this reason one should be careful using the codifference and the LCF in the range $\theta X_t \gg 1$. }\label{fig:LCF}
\end{figure}

If we want to analyse the codifference of the CIR model, it would be required to study the memory of the velocity $V_t = \sqrt{D_t} \dd B_t/\dd t$. But the white noise $\dd B_t/\dd t$ is not well-defined in a classical sense. It can be interpreted as a distribution which leads to a similar redefinition of the covariance, the familiar Dirac delta. The codifference is, however, non-linear and this approach fails. The solution is to consider only the well-defined velocity processes $V_t = \sqrt{D_t} Y_t$ with $Y_t$ being some classical process which models the velocity as being undisturbed by the fluctuations of the diffusivity. The behaviour of the white noise can be studied if we consider $t$ large enough such that $r_Y(t)=0$ strictly or approximately. It is natural to assume that $Y_t$ is Gaussian, while choosing the model of $D_t$ is more subtle.

 The CIR process for $ab\in \mathbb N$, can be proved to be a sum of squared independent Ornstein-Uhlenbeck processes, which follows directly from writing the stochastic differential equation of such a sum \cite{finMeth}. Thus, a natural generalisation is to consider $D_t$ being a square of a Gaussian process \cite{Chechkin2017,Sposini18}. We will assume that the velocity can be decomposed as
\bgeq
  V_t=\sigma |Z_t|Y_t,
\eeq
where both $Z_t$ and $Y_t$ are Gaussian with variance one. In this model we have ample freedom in describing a wide range of memory types, because any covariance $r_Z$ and $r_Y$ can be used. By choosing $r_Y$ we model the internal dynamics, if $r_Y(t)=0$ in the considered time scale we arrive back at \Ref{eq:dXDef}; by choosing $r_Z$ we model the memory structure of $D_t$: exponential, power law, oscillating, etc. The one-dimensional distributions are more rigged, as we limit ourselves to $D_t$ having the PDF of a square Gaussian, that is $\chi^2_1$ distribution (a special case of the gamma distribution). A rather technical derivation (Proposition \ref{prp:prod})  then shows that the exact form of the codifference is
\bgeq\label{eq:diffCodEx}
\tau_V^\theta(t)=\f{1}{\theta^2}\ln\lt(\f{\sqrt{1-r_Z(t)^2}}{\pi}\f{1+(\theta\sigma)^2}{(\theta\sigma)^2(1-r_Z(t)^2)+1}\sum_{\rho\in\{\rho_+,\rho_-\}}\f{\f{\pi}{2}+\mathrm{arctan}\lt(\f{\rho}{\sqrt{1-\rho^2}}\rt)}{\sqrt{1-\rho^2}}\rt)
\eeq
where
$$ \rho_\pm \defeq \f{ (\theta\sigma)^2(1-r_Z(t)^2)r_Y(t)\pm r_Z(t)}{(\theta\sigma)^2(1-r_Z(t)^2)+1}.$$
This formula looks complicated, but is composed only of elementary functions. It is illustrated in figure \ref{fig:diffCod}, were we plotted the codifference $\tau_V^\theta$ as a function of $r_Z$ and $r_Y$ for four different $\theta$s. Having calculated the codifference for at least two $\theta$s, one can solve the system of equations resulting from \Ref{eq:diffCodEx} and calculate $r_Z, r_Y$. This procedure may be considered simpler than using the covariance $r_Z$, which requires calculating the average of $|Z_sZ_{s+t}|$ given by a hard-to-evaluate integral. The covariance $r_V$ can also be obtained from taking the limit $\theta\to 0$ of the codifference.

\bg{figure}
\centering
\includegraphics[width=12cm]{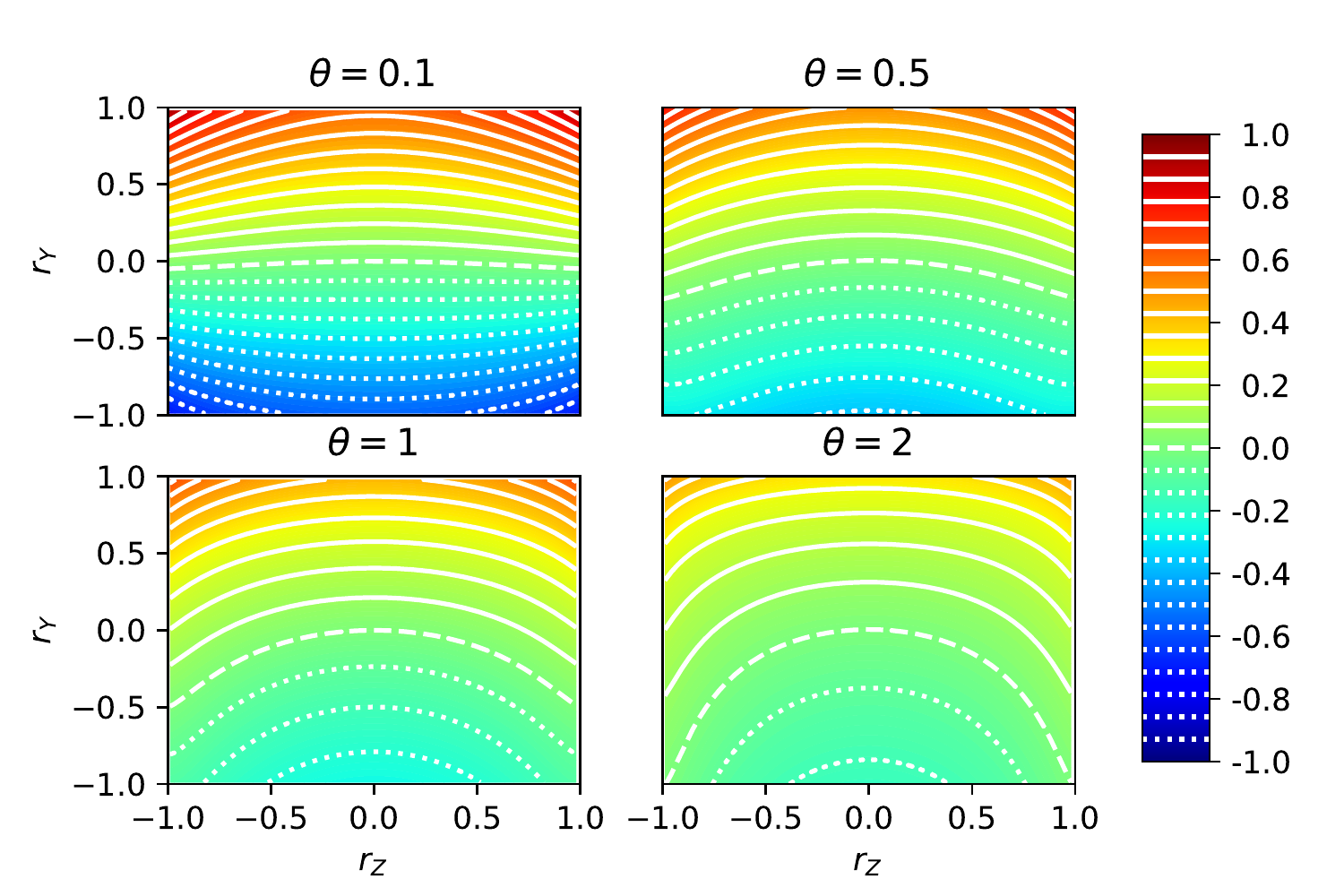}
\caption{Codifference $\tau_V^\theta$ as a function of $r_Z$ and $r_Y$ as given by Eq. \Ref{eq:diffCodEx}. White isolines are drawn at levels $\{\ldots,-2/14,-1/14,0,1/14,2/14,\ldots\}$. The dependence on $r_Z$ is symmetric, which can be seen directly from the definition $V_t=\sigma|Z_t|Y_t$. For larger $\theta$ the codifference varies less and the influence of the positive dependence of $D_t$ becomes dominating (the isolines become more concave). For a given $\tau_V^\theta$ the covariances $r_Z$ and $r_Y$ can be determined by looking for the crossing points of the corresponding isolines for at least 2 different values of $\theta$.}\label{fig:diffCod}
\end{figure}

More importantly, when $r_Y(t)=0$ the codifference is clearly non-zero, so it detects the dependence introduced by $D_t=Z_t^2$. Its asymptotic for small $r_Z(t)$ (e.g., at long times) in this case is the simple relation
\bgeq\label{eq:diffDiffAs1}
\tau_V^\theta(t) \sim \f{\sigma^2}{2(1+(\theta\sigma)^2)^2}r_Z(t)^2, \quad r_Z(t)\to 0.
\eeq
Thus the codifference detects the memory structure of the time-varying diffusion coefficient $D_t = Z_t^2$ even in the regime $r_Y(t)= 0$ in which the covariance $r_V(t)$ is zero and does not contain any important information. This is also true when $r_Z(t)=0$ but $r_Y(t)\neq 0$, this time the codifference is asymptotically proportional to $r_Y(t)$; the proportionality constant depends only on the one-dimensional distributions of $D$, the exact form of the dynamics does not matter, see Proposition \ref{prp:indepD}.

For some systems different models of $D_t$ may be more suitable. When $D_t$ is strongly concentrated around its mean value a possible choice is a simple Gaussian centred around some $b$, $V_t = (\sigma Z_t+b)Y_t$. This model permits the unphysical situation when $D_t<0$, but when $\sigma\ll b$ the probability of this event is negligible. In this case an elementary formula for the codifference also can be given (see \Ref{eq:ZYa}) and again even for $r_Y(t)=0$ the internal dependence of $D_t$ is still detected, this time with asymptotic
\bgeq\label{eq:diffDiffAs2}
\tau_V^\theta(t)\sim \f{b^2}{(1+(\theta\sigma)^2)^2}r_Z(t),\quad r_Z(t)\to 0.
\eeq

\subsection{Discussion}

The aim of this work was to provide the theoretical background for using
the codifference as a dependence measure suited for the study of various
non-Gaussian and ergodicity breaking models. This goal was achieved in
few steps. First we proved that the codifference has intuitive properties
that one would expect from a reasonable memory function, such as additivity,
positivity for the case of complete dependence and being null for the case
of independence. Second,
we showed that it can be calculated using fairly straightforward methods for
typical random parameters and diffusing-diffusivity models, which represent
a significant extension of the previously established results for stable and
infinitely divisible processes. Finally, we analysed how the codifference
detects forms of dependence and ergodicity breaking which cannot be easily
studied using solely covariance-based methods.

We also showed one example of non-detected ergodicity breaking, the case of
a Langevin equation with a random return rate. In this case we offer an
easy fix: the codifference works well for the increments of this process.
We note that within this paper we did not analyse ergodicity breaking caused
by ageing. In principle, the codifference should work, but the analytical
analysis will be challenging for many of these phenomena.

In addition to the codifference, we also discussed a related quantity, the
logarithm of the characteristic function (LCF), which was interpreted as a
measure of dispersion. Our contribution is an extension of the Fourier methods
and a distinct view based on ideas previously developed only for heavy tailed
$\alpha$-stable distributions. The codifference is also very closely related
to the theory of the dynamical functional, which was already successfully
used for real data, and should be considered a part of the same framework.

The cost of using this technique is that linearity is a powerful analytical
tool, especially for complicated models, and a significant part of this strength
is lost when using the codifference. The more complicated defining formula also
may make its form more complicated (e.g., see Table \ref{tab:cdf}). However,
it is a clear application of the characteristic function which does not seem to be
commonly acknowledged and the Fourier-based techniques by themselves are widely
used by the scientific community. Thus, it has an advantage,  offering a wide
choice of established analytical methods and estimation techniques. In some
cases (e.g., \Ref{eq:diffCodEx}) the codifference has a simpler form than the
covariance.

We believe that the most important example that was considered was also
the simplest: deterministic motion with its scale (diffusion coefficient)
varying from trajectory to trajectory. The observed asymptotical behaviour
of the codifference contains a lot of useful information and lays the foundation for
possible future applications in more complex and realistic models, some of
which we discussed. At the same time we stress that even this initial, highly
simplified model is being commonly used, especially in biophysical systems.

We are confident that the obtained results are interesting in their own right,
but we also promote their additional value by indicating the limitations
of the methodology based on the MSD and the covariance. Both are, without a
doubt, essential parts of the scientific language related to diffusion and
complex phenomena, but their limitations are becoming more and more evident,
as contemporary research starts to concentrate around non-Gaussian systems
with complicated memory structure; the change is stimulated by increasing
experimental evidence. These complex and non-linear phenomena require new
complex and non-linear methods.

\setcounter{equation}{0}

\section{Derivations}

\subsection{Basic definitions and properties}

All processes considered in this work can be labelled as "conditionally Gaussian". In practical applications these processes are Gaussian locally, in the temporal or spatial sense. The formal definition is more general. 
\bg{dfn} We call a process conditionally Gaussian when any of its finite-dimensional distributions is a Gaussian distribution under some conditioning by $\sigma$-algebra $\mathcal{C}$. That is, any finite dimensional distribution $\bd X \defeq [X_{t_1},\ldots, X_{t_n}]$ can be written as
\bgeq
\bd X = A \bd Y + \bd \mu,
\eeq
where $A$ and $ \bd \mu$ are a $\mathcal C$-measurable $n\times n$ random matrix and an $n$-dimensional random vector. Both may depend on $t_1,\ldots, t_n$. The vector $\bd Y$ is i.i.d $\mathcal N(0,1)$ and is independent of $A$ and $\bd \mu$.

If $\bd \mu = 0$ for any $t_1,\ldots,t_n$ we call a process conditionally centred Gaussian. Further on we will consider only this class. Similarly, we call a process conditionally stationary Gaussian, if the distribution of  $A$ and $\bd\mu$ does not depend on time translation $t_1,\ldots, t_n\mapsto t_1+t,\ldots,t_n+t$.
\end{dfn}
\bg{prp}
The distribution of a conditionally Gaussian process is completely determined by the knowledge of $\mathcal C$, the conditional mean and the conditional covariance
\bgeq\label{eq:covDefFull}
\mu_X(t|\mathcal C) = \E[X_t|\mathcal C],\quad r_X(s,t|\mathcal C)\defeq \E[X_sX_t|\mathcal C].
\eeq
The process is conditionally centred if and only if $\mu_X(t|\mathcal C) = 0$. The process is conditionally stationary if and only if $\mu_X(t|\mathcal C)=\const$ and $r_X(s,t|\mathcal C)$ is a function of $t-s$, denoted $r_X(t-s|\mathcal C)$.
\end{prp}
\bg{proof} This is a direct consequence of the equality
\bgeq
\mathrm{P}(X_{t_1}\in A_1,\ldots, X_{t_n}\in A_n) = \EE{ \mathrm{P}(X_{t_1}\in A_1,\ldots, X_{t_n}\in A_n|\mathcal C)}.
\eeq
The conditional probability on the right is a Gaussian integral and a function  of $\mu_X(t|\mathcal C)$ and $r_X(s,t|\mathcal C)$. The representation of conditionally centred and stationary processes are just a reflection of the analogical representations for Gaussian processes.
\end{proof}

\bg{dfn}\label{dfn:codiff} We define the codifference function as
\bgeq\label{eq:codDefFull}
\tau_X^\theta(s,t)\defeq \f{1}{\theta^2}\ln\f{\E\lt[\e^{\I\theta( X_t- X_s)}\rt]}{\E\lt[\e^{\I\theta X_t}\rt]\E\lt[\e^{-\I\theta X_s}\rt]}.
\eeq
For stationary process it is a function of $t-s$, which we denote as $\tau_X^\theta(t)$, similarly as for the covariance, see also Eq. \Ref{eq:cdfDef}.

Additionally, we define the log characteristic function (LCF) as
\bgeq
\zeta_X^\theta(t)\defeq -\f{2}{\theta^2}\ln\EE{\e^{\I\theta(X_t-\mu_X(t))}}.
\eeq
\end{dfn}
All expected values in the above definitions are finite, but they may be complex and the denominator may be 0. This is however not the case in the class of processes considered herein.

\bg{prp}\label{prp:basic} For any conditionally centred Gaussian process the codifference and the LCF are well-defined real-valued functions.
\end{prp}

\bg{proof}
The Gaussian function centred at 0 is positive-definite. The mixture of positive-definite functions is positive-definite. Therefore all expected values in Definition \ref{dfn:codiff} are real numbers larger than 0 and less or equal 1. The logarithms are therefore real.
\end{proof}
We also note that for conditionally centred Gaussian processes a reduced formula for the codifference is available,
\begin{eqnarray}
\nonumber
\label{eq:tauCexpl}
\tau_X^\theta(s,t)&=&\f{1}{\theta^2}\ln\f{\EE{\EE{\e^{\I\theta(X_t-X_s)}|\mathcal C}}}{ \EE{\EE{\e^{\I\theta X_t}|\mathcal C}}\EE{\EE{\e^{-\I\theta X_s}|\mathcal C}}}\\
&=&\f{1}{\theta^2}\ln\f{\EE{\e^{-\theta^2(r_X(t,t|\mathcal C)-2r_X(s,t|\mathcal C)+r_X(s,s|\mathcal C))/2}}}{\EE{\e^{-\theta^2 r_X(t,t|\mathcal C)/2}}\EE{\e^{-\theta^2r_X(s,s|\mathcal C)/2}}},
\end{eqnarray}
which is very useful for calculations. For non-centred process the additional term
\bgeq
+\f{1}{\theta^2}\ln \f{\EE{\e^{\I\theta (\mu_X(t|\mathcal C)-\mu_X(s|\mathcal C))}}}{\EE{\e^{\I\theta \mu_X(t|\mathcal C)}}\EE{\e^{-\I\theta \mu_X(s|\mathcal C)}}}
\eeq
appears.
Here all averages are finite, but they can generally be complex values, moreover in particular cases the averages in the denominator can be 0. This strongly suggests the codifference should be used carefully in this case (the same applies to the LCF).

Additionally, representation \Ref{eq:tauCexpl} yields another  desirable property of the codifference:
\bg{prp}\label{prp:codPos} For a conditionally centred Gaussian process with positive covariance $r_X(s,t|\mathcal C)$ the codifference $\tau_X^\theta(s,t)$ is also positive, a negative conditional covariance implies negative codifference.
\end{prp}
If the support of $r_X(s,t|\mathcal C)$ is on both positive and negative half-axes, the sign of the codifference may vary, but it is worth noting that with $r_X(t,t|\mathcal C)$ and $r_X(s,s|\mathcal C)$ fixed, it depends monotonically on $r_X(s,t|\mathcal C)$, so if the conditional covariance is smaller in the sense of stochastic dominance, the codifference will also be smaller.

Now, a simple fact follows only from the expansion $\ln(x)\in x-1+ o(x)$ as $ x\to 1$.
\bg{prp}
For any stationary process $X$ with asymptotically independent values
\bgeq
\tau_X^\theta(t)\sim \f{1}{\theta^2}\lt(\f{\E\lt[\e^{\I\theta (X_{s+t}-X_s)}\rt]}{\E\lt[\e^{\I \theta X_{s+t}}\rt]\E\lt[\e^{-\I\theta X_s}\rt]}-1\rt),\quad t\to\infty.
\eeq
\end{prp}
\bg{proof}
We assume that $X_{s+t}$ and $X_s$ are asymptotically independent as $t\to\infty$ (note that this property is not sufficient to imply that $X$ is mixing). Therefore
\bgeq
\E\lt[\e^{\I\theta(X_{s+t}-X_s)}\rt]\xrightarrow{t\to\infty} \E\lt[\e^{\I\theta X_{s+t}}\rt]\E\lt[\e^{-\I\theta X_s}\rt],
\eeq
and the ratio of expected values under the logarithm converges to 1 so we can use the expansion $\ln(x)\approx x-1$.
\end{proof}
This simple fact is a prototype for the later results, which describe cases when it is possible to remove the non-linear logarithmic function if the process can be somehow decomposed as a transformation of some weakly dependent variables.

If the process $X$ does not have asymptotically independent values the non-linearity cannot be removed at $t\to\infty$, but if it is an ensemble of such processes (i.e., the conditioned process is mixing), it can be shown that the codifference converges to a positive constant, non-linearly dependent on the law of $D$.

\subsection{Random parameter models}

\bg{prp}\label{prp:codInf}
If the process $X$ is an ensemble of mixing stationary centred Gaussian processes, then, denoting $D = \EE{X_t^2|\mathcal C}$,
\bgeq
\tau_X^\theta(\infty)=\f{1}{\theta^2}\ln\f{\E\lt[\e^{-\theta^2 D}\rt]}{\E\lt[\e^{-\theta^2 D/2}\rt]^2}\ge 0
\eeq 
and equal 0 only for deterministic $D$.
\end{prp}
\bg{proof} The calculation is simple. Because $r_X(t|\mathcal C)\le D$ almost surely the random variable $\e^{\theta^2(r_X(t|\mathcal C)-D)}$ is positive and bounded by 1 for every $t$. We can commute the  limit with the logarithm and the averaging, getting
\bgeq
\lim_{t\to\infty}\tau_X^\theta(t)=\f{1}{\theta^2}\ln\f{\E\lt[\lim_{t\to\infty}\e^{\theta^2(r_X(t|\mathcal C)-D)}\rt]}{\E\lt[\e^{-\theta^2D/2}\rt]}=\f{1}{\theta^2}\ln\f{\E\lt[\e^{-\theta^2D}\rt]}{\E\lt[\e^{-\theta^2D/2}\rt]^2}.
\eeq
The non-negativity of the above stems from Jensen's inequality applied to the function $x\mapsto x^2$ and the variable $\e^{-\theta^2D/2}$. 

\end{proof}
\noindent\textbf{Remark.} A similar calculation repeated for symmetrised codifference \Ref{eq:symmCodDef} shows that it does not exhibit this behaviour. Under the same assumptions
\begin{align}\label{eq:symmCod0}
\widetilde \tau_X^\theta(\infty) &= \lim_{t\to\infty}\f{1}{2\theta^2} \ln\f{\E\lt[\e^{\I\theta(X_{s+t}-X_s)}\rt]}{\E\lt[\e^{\I\theta(X_{s+t}+X_s)}\rt]}=\f{1}{2\theta^2}\ln\f{\E\lt[\lim_{t\to\infty}\e^{\theta^2(r_X(t|\mathcal C)-D)}\rt]}{\E\lt[\lim_{t\to\infty}\e^{\theta^2(-r_X(t|\mathcal C)-D)}\rt]}
\nonumber\\
&= \f{1}{\theta^2}\ln\f{\E\lt[\e^{-\theta^2D}\rt]}{\E\lt[\e^{-\theta^2D}\rt]} = 0,
\end{align}
i.e., it cannot detect this form of residual dependence and ergodicity breaking.

\bg{prp}\label{prp:randD}
Let the process $X$ have the form
\bgeq
X_t=\sqrt{D}Y_t,
\eeq
where $Y$ is a stationary Gaussian process, $\E[Y_t^2]=1$, and $D>0$ is a random variable independent of $Y$. Then the codifference has the form
\bgeq\label{eq:randD}
\tau_X^\theta(t) = \f{1}{\theta^2}\ln\f{\E\lt[\e^{-\theta^2D(1-r_Y(t))}\rt]}{\E\lt[\e^{-\theta^2D/2}\rt]^2}.
\eeq

\bg{itemize}
\item[a)] It is additive with respect to $D$, that is if $D=D'+D''$ for independent $D'$ and $D''$, then
\bgeq
\tau_X^\theta(t) = \tau_{X'}^\theta(t)+\tau_{X''}^\theta(t)
\eeq
where $X'_t=\sqrt{D'}Y_t$ and $X''_t=\sqrt{D''}Y_t$.

\item[b)] It is an increasing function of the covariance $r_Y(t)$, which is smaller than $r_X(t)$ for $r_Y(t)$ close to $1$ and larger than $r_X(t)$ when the latter is close to 0. If $\E[D]<\infty$ the difference $\tau_X^\theta(t)-r_X(t)$ decreases as a function of $r_Y(t)$. 
\item[c)] For any mixing $Y$ the difference $\tau_X^\theta(t)-\tau_X^\theta(\infty)$ exhibits the same type of asymptotic as the covariance $r_Y(t)$, that is
\bgeq
\tau_X^\theta(t)-\tau_X^\theta(\infty)\sim \f{\E\lt[D\e^{-\theta^2 D}\rt]}{\E\lt[\e^{-\theta^2 D}\rt]}r_Y(t),\quad t\to\infty.
\eeq
\end{itemize}
\end{prp}
\bg{proof}
Let us start from writing the conditional covariance,
\bgeq
\E[X_t^2|D]=D,\quad \E[(X_{s+t}-X_s)^2|D] = 2D(1-r_Y(t)),
\eeq
which implies that
\bgeq
\tau_X^\theta(t) = \f{1}{\theta^2}\ln\f{\E\lt[\e^{-\theta^2D(1-r_Y(t))}\rt]}{\E\lt[\e^{-\theta^2D/2}\rt]^2}.
\eeq
If we substitute $D=D'+D''$ both numerator and denominator factorise as products of independent random variables. The formula
\bgeq
\tau_X^\theta(t)=\tau_{X'}^\theta(t)+\tau_{X''}^\theta(t)
\eeq
follows.

In point b) the monotonic dependence is a consequence of the fact that only the numerator of the fraction in \Ref{eq:randD} depends on $r_Y(t)$. It is a Laplace transform of the variable $D$ calculated at the point $\theta^2(1-r_Y(t))$, it decreases as the argument increases, so it is an increasing function of $r_Y(t)$. This dependence is continuous. When $r_X(t)=0$, e.g., always for $t=0$ formula \Ref{eq:randD} simplifies and we can apply Jensen's inequality,
\bgeq
\tau_X^\theta(0)=-\f{2}{\theta^2}\ln\E\lt[\e^{-\theta^2D/2}\rt]\le -\f{2}{\theta^2}\E\lt[\ln\e^{-\theta^2D/2}\rt] = \E[D] = r_X(0).
\eeq
For $r_Y(t)$ close to 0 we can use Proposition \ref{prp:codInf} to determine, that the codifference is positive. For the last property listed in b), let us write the difference $\tau_X^\theta(t)-r_X^\theta(t)$ as a function of $r=r_X\theta(t)$,

\bgeq
f(r)\defeq  \f{1}{\theta^2}\ln\f{\E\lt[\e^{-\theta^2D(1-r))}\rt]}{\E\lt[\e^{-\theta^2D/2}\rt]^2}-\E[D]r = \f{1}{\theta^2}\ln\f{\E\lt[\e^{-\theta^2(D-\E[D])(1-r))}\rt]}{\E\lt[\e^{-\theta^2(D-\E[D])/2}\rt]^2}.
\eeq
Using the majorised convergence theorem, the derivative of the numerator exists and determines the sign of $f'$. Denoting $F_r\defeq \theta^2(D-\E[D])(1-r)$ we have
\bgeq
f'(r)\propto \f{1}{1-r}\EE{F_r\e^{-F_r}} = \f{1}{1-r}\EE{F_r(\e^{-F_r}-1)}\le 0,
\eeq
where we used the fact that $\E[F_r]=0$ and $x(\e^{-x}-1)\le 0$.

For c)  consider $\tau_X(t)-\tau_X(\infty)$ and use the expansion $\ln(x)\approx x-1$
\bgeq
\tau_X^\theta(t)-\tau_X^\theta(\infty)=\f{1}{\theta^2}\ln\f{\E\lt[\e^{-\theta^2D(1-r_Y(t))}\rt]}{\E\lt[\e^{-\theta^2D}\rt]}\sim\f{1}{\theta^2}\lt( \f{\E\lt[\e^{-\theta^2D(1-r_Y(t))}\rt]}{\E\lt[\e^{-\theta^2D}\rt]}-1\rt),\quad t\to \infty.
\eeq
Now we can rearrange the right side of the above equation and get
\bgeq
\lim_{t\to\infty}\f{\tau_X^\theta(t)-\tau_X^\theta(\infty)}{r_Y(t)}=\lim_{t\to\infty}\f{\E\lt[\f{\e^{\theta^2Dr_Y(t)}-1}{\theta^2r_Y(t)}\e^{-\theta^2D}\rt]}{\E\lt[\e^{-\theta^2D}\rt]}=\lim_{x\to 0}\f{\E\lt[\f{\e^{Dx}-1}{x}\e^{-\theta^2D}\rt]}{\E\lt[\e^{-\theta^2D}\rt]} = \f{\E\lt[D\e^{-\theta^2D}\rt]}{\E\lt[\e^{-\theta^2D}\rt]}.
\eeq
\end{proof}
The analogues of a) and b) also hold for the LCF, the derivation is very similar as in Proposition \ref{prp:randD} so we only state the result.
\bg{prp}\label{prp:randDLCF}
Let the process $X$ have the form
\bgeq
X_t=\sqrt{D}Y_t,
\eeq
where $Y$ is a centred Gaussian process and $D>0$ is a random variable independent of $Y$. 

Then the LCF has the form
\bgeq\label{eq:randDLCF}
\zeta_X^\theta(t) = -\f{2}{\theta^2}\ln\E\lt[\e^{-\theta^2D\delta_Y^2(t)/2}\rt]
\eeq
and:
\bg{itemize}
\item[a)] If $\E[D]<\infty$ then
\bgeq
\zeta_X^\theta (t)\sim \delta_X^2(t),\quad t\to 0^+.
\eeq
\item[b)] It is additive with respect to $D$, that is if $D=D'+D''$ for independent $D'$ and $D''$, then
\bgeq
\zeta_X^\theta(t) = \zeta_{X'}^\theta(t)+\zeta_{X''}^\theta(t)
\eeq
where $X'_t=\sqrt{D'}Y_t$ and $X''_t=\sqrt{D''}Y_t$.
\item[c)] It is an increasing function of the MSD $\delta_Y^2(t)$.
\item[d)] For $\E[D]<\infty$ the difference $\delta_X^2(t)-\zeta_X^\theta(t)$ is non-negative and increases as $\delta_X^2(t)$ increases.
\end{itemize}
\end{prp}

The asymptotic of the codifference near zero depends on the tail behaviour of $p_D$ and can be used to study it. This statement is clarified by the following result.

\bg{prp}\label{prp:codShortTime}
If the stationary Gaussian process $Y$ is mean-square continuous and $X_t=\sqrt{D}Y_t$, then
\bg{itemize}
\item[a)] for $\E[D]<\infty$ 
\bgeq
\tau_X^\theta(0)-\tau_X^\theta(t)\sim \E[D](1-r_Y(t)),\quad t\to 0^+
\eeq
and
\bgeq
\zeta_X^\theta(t) \sim \E[D]\delta_Y^2(t),\quad t\to 0^+.
\eeq
\item[b)] If
\bgeq
p_D(d)\sim  \f{L(d)}{d^{1+\rho}},\quad 0<\rho<1, \quad d\to\infty 
\eeq
for some slowly varying function $L$, then
\bgeq
\tau_X^\theta(0)-\tau_X^\theta(t)\sim \theta^{2\rho-2}\f{\Gamma(1-\rho)}{\rho} L(\theta^{-2}(1-r_Y(t))^{-1})(1-r_Y(t))^\rho,\quad t\to 0^+.
\eeq
\end{itemize}
\end{prp}

\bg{proof}
For a mean-square continuous $Y$ the covariance $r_Y$ is a continuous function. The codifference is also continuous and $\ln(x)\approx x-1$ implies that
\bgeq
\tau_X^\theta(0)-\tau_X^\theta(t)=-\f{1}{\theta^2}\ln\E\lt[\e^{-\theta^2D(1-r_Y(t))}\rt]\sim \f{1}{\theta^2}\lt(1-\E\lt[\e^{-D(1-r_Y(t))}\rt]\rt),\quad t\to 0^+.
\eeq
Because
\bgeq
\lim_{t\to 0^+}\E\lt[\f{1-\e^{-\theta^2D(1-r_Y(t))}}{\theta^2(1-r_Y(t))}\rt]=\lim_{x\to 0^+}\E\lt[\f{1-\e^{-Dx}}{x}\rt] = \E[D].
\eeq
The derivation for $\zeta_X^\theta$ is similar. For point b) we write the asymptotic of $\tau_X^\theta(0)-\tau_X^\theta(t)$ as the integral
\bgeq
\tau_X^\theta(0)-\tau_X^\theta(t)\sim \f{1}{\theta^2}\int_0^\infty\dd d\ \lt(1-\e^{-\theta^2d(1-r_Y(t))}\rt)p_D(d)
\eeq
and simplify the ratio under investigation
\begin{align}
&\f{\tau_X^\theta(0)-\tau_X^\theta(t)}{L(\theta^{-2}(1-r_Y(t))^{-1})(1-r_Y(t))^\rho}\sim\theta^{2\rho-2} \lim_{x\to 0^{+}}\int_0^\infty\dd d\ \lt(1-\e^{-xd}\rt)\f{1}{L(x^{-1})x^{\rho}}p_D(d) \nonumber\\
&= \theta^{2\rho-2} \int_0^\infty\dd d\ \f{1-\e^{-d}}{d} d^{-\rho} \lim_{x\to 0^+}\lt(\f{d}{x}\rt)^{\rho+1}\f{1}{L(x^{-1})}p_D\lt(\f{d}{x}\rt)\nonumber\\
&=\theta^{2\rho-2}\int_0^\infty\dd d\ \f{1-\e^{-d}}{d} d^{-\rho}=-\theta^{2\rho-2}\Gamma(-\rho)=\theta^{2\rho-2}\f{\Gamma(1-\rho)}{\rho}
\end{align}
\end{proof}

Now, let us move our attention from a random $D$ to the class of processes, for which the shape of the covariance function varies from trajectory to trajectory:

\bg{prp}\label{prp:fixVar}
For a mixture of stationary Gaussian processes with fixed non-random scale $D=\sigma^2$

\bgeq
\tau_X^\theta(t)=\f{1}{\theta^2}\ln \E\lt[\e^{\theta^2r_X(t|\mathcal C)}\rt].
\eeq
The above formula also implies that 

\bgeq
r_X(t)\le \tau_X^\theta(t)\le\f{1}{\theta^2}\lt((\theta\sigma)^{-2}\sinh(\theta^2\sigma^2) r_X(t)+\cosh(\theta^2\sigma^2)-1\rt).
\eeq
\end{prp}
\bg{proof}
Assumption of a fixed variance means that $\E[X(t)^2|\mathcal C] =\sigma^2$ for some deterministic $\sigma^2$. Using the conditional expectancy it follows that
\begin{eqnarray}
\nonumber
\tau_X^\theta(t)&=&\f{1}{\theta^2}\ln\f{\E\lt[\E\lt[\e^{\I\theta(X_{s+t}-X_s)}|\mathcal C\rt]\rt]}{\E\lt[\E\lt[\e^{\I\theta X_{s+t}}|\mathcal C\rt]\rt]\E\lt[\E\lt[\e^{-\I X_s}|\mathcal C\rt]\rt]} = \f{1}{\theta^2}\ln\f{\E\lt[\e^{-\theta^2(\sigma^2-r_X(t|\mathcal C))}\rt] }{\lt(\e^{-\theta^2\sigma^2/2}\rt)^2}\\
&=&\f{1}{\theta^2}\ln \E\lt[\e^{\theta^2r_X(t|\mathcal C)}\rt].
\end{eqnarray}
Now the left inequality is just Jensen's inequality applied to the function $\ln$. The right inequality follows from two approximations: the first is  $\ln x \le x-1$, the second is $\exp(x)\le L^{-1}\sinh(L)x+\cosh(L)$ for $-L\le x \le L$.
\end{proof}

For the exponentially decaying conditional covariance stronger results are available:

\bg{prp}\label{prp:randLang} For a mixture of stationary centred Gaussian processes with conditional covariance $r_X(t|\Lambda,D)=D\e^{-t\Lambda}$, with $\Lambda$ and $D$ independent, we observe the following asymptotic properties.
\bg{itemize}
\item[a)] Power law behaviour: if  $p_\Lambda(\lambda)\sim L(\lambda)\lambda^{\alpha-1},\lambda\to 0^+$ for slowly varying $L$, then

\bgeq\label{eq:langAsympt}
\tau_X^\theta(t)-\tau_X^\theta(\infty)\sim  C_{\alpha,\theta} \f{L(t^{-1})}{t^\alpha},
\eeq
where the constant $C_{\alpha,\theta}$ is
\bgeq
C_{\alpha,\theta}=\f{\Gamma(\alpha) }{\theta^2\E\lt[\e^{-\theta^2D}\rt]^2}\sum_{k=1}^\infty\theta^{2k}\f{\E\lt[D^k\e^{-\theta^2D}\rt]}{k!}\f{1}{k^\alpha}.
\eeq
\item[b)] Quick decay behaviour: if  $p_\Lambda(\lambda) \in \mathcal O(\lambda^\infty),\lambda\to 0^+$ then
\bgeq
\tau_X^\theta(t)-\tau_X^\theta(\infty)\in\mathcal O(t^{-\infty}),\quad t\to\infty.
\eeq
\item[c)] Truncation: if $\Lambda=\lambda_0+\widetilde \Lambda$ for deterministic $\lambda_0>0$ then
\bgeq
\tau_X^\theta(t)-\tau_X^\theta(\infty)\le \e^{-\lambda_0 t}\lt(\tau_{\widetilde X}^\theta(t)-\tau_{\widetilde{X}}^\theta(\infty)\rt),
\eeq
where $\widetilde X$ is a solution of the Langevin equation with viscosity $\widetilde \Lambda$ and the same $D$.
\end{itemize}
\end{prp}
\bg{proof}

For a) first we apply the expansion $\ln(x)\approx x-1$ to $\tau_X^\theta(t)-\tau_X^\theta(\infty)$
\bgeq
\tau_X^\theta(t)-\tau_X^\theta(\infty)=\f{1}{\theta^2}\ln\f{\E\lt[\e^{-\theta^2D\lt(1-\e^{-t\Lambda}\rt)}\rt]}{\E\lt[\e^{-\theta^2D}\rt]}\sim\f{1}{\theta^2}\lt( \f{\E\lt[\e^{-\theta^2D\lt(1-\e^{-t\Lambda}\rt)}\rt]}{\E\lt[\e^{-\theta^2D}\rt]} -1\rt),\quad t\to\infty.
\eeq
Therefore
\begin{align}
\label{eq:langSum}
\nonumber
\tau_X^\theta(t)-\tau_X^\theta(\infty)&\sim\f{1}{\theta^{2}\E\lt[\e^{-\theta^2D}\rt]}\E\lt[\e^{-\theta^2D}\lt(\e^{\theta^2D\e^{-t\Lambda}}-1\rt)\rt]\\
\nonumber
&=\f{1}{\theta^2\E\lt[\e^{-\theta^2D}\rt]}\E\lt[\e^{-\theta^2D}\sum_{k=1}^\infty \theta^{2k}\f{D^k}{k!}\e^{-kt\Lambda}\rt]\nonumber\\
&=\f{1}{\theta^2\E\lt[\e^{-\theta^2D}\rt]}\sum_{k=1}^\infty \theta^{2k}\f{\E\lt[D^k\e^{-\theta^2D}\rt]}{k!}\E\lt[\e^{-kt\Lambda}\rt],\quad t\to \infty.
\end{align}

Note that the sum within consists of positive terms, so the commutation of expectation and sum is justified.

Now, knowing the asymptotic $p_\Lambda(\lambda)\sim \lambda^{\alpha-1},\lambda\to 0^+$ we can apply the Tauberian theorem 
\bgeq
\f{t^\alpha}{L(t^{-1})} \E\lt[\e^{-kt\Lambda}\rt]\xrightarrow{t\to\infty} \Gamma(\alpha) \f{1}{k^\alpha}.
\eeq
The sum \Ref{eq:langSum} consists of positive terms, so let us study its asymptotic
\begin{align}
&\f{t^\alpha}{L(t^{-1})}\sum_{k=1}^\infty\theta^{2k}\f{\E\lt[D^k\e^{-\theta^2D}\rt]}{k!}\E\lt[\e^{-kt\Lambda}\rt]=\sum_{k=1}^\infty\theta^{2k}\f{\E\lt[D^k\e^{-\theta^2D}\rt]}{k!}\f{t^\alpha}{L(t^{-1})} \E\lt[\e^{-kt\Lambda}\rt]\nonumber\\
&\xrightarrow{t\to\infty} \Gamma(\alpha)\sum_{k=1}^\infty\theta^{2k}\f{\E\lt[D^k\e^{-\theta^2D}\rt]}{k!}\f{1}{k^\alpha},
\end{align}
where the commutation of taking the limit and the sum is justified by the inequality
\bgeq
t^\alpha\E\lt[\e^{-k t \Lambda}\rt]\le t^\alpha\E\lt[\e^{-t \Lambda}\rt].
\eeq 
The right term is convergent with respect to $t$, therefore it is bounded, so the left term is uniformly bounded with respect to $k$ and we can use the dominated convergence theorem.

Note that the resulting sum is also bounded with respect to $\alpha$,
\begin{eqnarray}
\nonumber
\fl\sum_{k=1}^\infty\theta^{2k}\f{\E\lt[D^k\e^{-\theta^2D}\rt]}{k!}\f{1}{k^\alpha}&<&\sum_{k=1}^\infty\theta^{2k}\f{\E\lt[D^k\e^{-\theta^2D}\rt]}{k!}\\
\fl&=&\E\lt[\sum_{k=0}^\infty\theta^{2k} \f{D^k}{k!}\e^{-\theta^2D}\rt]-\E\lt[\e^{-\theta^2D}\rt] = 1-\E\lt[\e^{-\theta^2D}\rt].
\end{eqnarray}
This concludes the derivation of a). Now let us prove b). We fix integer $N>0$ and then make the estimation
\begin{align}
\E\lt[\e^{-kt\Lambda}\rt]t^{N}&=\int_0^\infty \dd \lambda\  p_\Lambda(\lambda) \e^{-kt\lambda} t^{N} = k^{N+1}\int_0^\infty \dd \lambda\   p_\Lambda\lt(\f{\lambda}{kt}\rt)\e^{-\lambda }\lambda^{N+1} \lt(\f{\lambda}{kt}\rt)^{-N-1}\nonumber\\
&\xrightarrow{t\to\infty} 0
\end{align}
to obtain
\bgeq
\lim_{t\to\infty}\lt(\tau_X^\theta(t)-\tau_X^\theta(\infty)\rt) t^{N}=\lim_{t\to\infty}\f{1}{\theta^2}\sum_{k=1}^\infty\theta^{2k} \f{\E\lt[D^k\e^{-\theta^2D}\rt]}{k!}\E\lt[\e^{-kt\Lambda}\rt]t^{N}=0.
\eeq
The limit follows because it is a convergent sum of positive terms.

In order to prove the last point c) notice that $\e^{-t\lambda_0}<1 $ and $ \e^{t\lambda_0}>1$ so $x\mapsto x^{\e^{-t\lambda_0}}$ is a concave function and $\e^{-\theta^2D\e^{t\lambda_0}}\le\e^{-\theta^2D}$. Therefore
\begin{align}
\tau_X^\theta(t)-\tau_X^\theta(\infty)&=\f{1}{\theta^2}\ln\f{\E\lt[\e^{-\theta^2D}\e^{\theta^2D\e^{-t\Lambda}\e^{-\lambda_0t}}\rt]}{\E\lt[\e^{-\theta^2D}\rt]} = \f{1}{\theta^2}\ln\f{\E\lt[\lt(\e^{-\theta^2D\e^{t\lambda_0}}\e^{\theta^2D\e^{-t\Lambda}}\rt)^{\e^{-t\lambda_0}}\rt]}{\E\lt[\e^{-\theta^2D}\rt]}\nonumber\\
&\le  \f{1}{\theta^2}\ln\f{\E\lt[\e^{-\theta^2D\e^{t\lambda_0}}\e^{\theta^2D\e^{-t\Lambda}}\rt]^{\e^{-t\lambda_0}}}{\E\lt[\e^{-\theta^2D}\rt]}=\e^{-t\lambda_0} \f{1}{\theta^2}\ln\f{\E\lt[\e^{-\theta^2D\e^{t\lambda_0}}\e^{\theta^2D\e^{-t\Lambda}}\rt]}{\E\lt[\e^{-\theta^2D}\rt]}\nonumber\\
&\le \e^{-t\lambda_0} \f{1}{\theta^2}\ln\f{\E\lt[\e^{-\theta^2D}\e^{\theta^2D\e^{-t\Lambda}}\rt]}{\E\lt[\e^{-\theta^2D}\rt]} =\e^{-t\lambda_0}\lt(\tau_{\widetilde X}^\theta(t)-\tau_{\widetilde X}^\theta(\infty)\rt).
\end{align}
\end{proof}

In the next proposition we will study the properties of the increment process
\bgeq
\Delta X_t\defeq X_{t+\Delta t}-X_t
\eeq
and use it to detect non-ergodicity. 
\bg{prp}\label{prp:increments}
Considering the same process as in Proposition \ref{prp:randLang}, the codifference of its increments $\Delta X_t$  converges to a constant
\bgeq
\lim_{t\to\infty}\tau_{\Delta X}^\theta(t) = \f{1}{\theta^2}\ln\f{\E\lt[\e^{2\theta^2D\lt(\e^{-\Delta t\Lambda}-1\rt)}\rt]}{\E\lt[\e^{\theta^2D\lt(\e^{-\Delta t\Lambda}-1\rt)}\rt]^2}\ge 0,
\eeq
which equals 0 only when both $D$ and $\Lambda$ are deterministic. After suitable rescaling $\Delta\widetilde X_t\defeq \Delta X_t/\sqrt{\Delta t}$ the limit becomes independent of $\Delta t$,
\bgeq
\lim_{\Delta t\to 0^+}\lim_{t\to\infty}\tau_{\Delta \widetilde X}^\theta(t)= \f{1}{\theta^2}\ln\f{\E\lt[\e^{-2\theta^2D\Lambda}\rt]}{\E\lt[\e^{-\theta^2D\Lambda}\rt]^2}.
\eeq
\end{prp}
\bg{proof}
The reasoning is similar to the one  shown in the proof of Proposition \ref{prp:randD} b). The increment process $\Delta X_t$ is a stationary process, which is conditionally Gaussian. We can calculate its conditional variance
\bgeq
\E[\Delta X_t^2|\Lambda,D] = 2D\lt(1- \e^{-\Delta t\Lambda}\rt)
\eeq
and the variance of the difference
\begin{eqnarray}
\nonumber
\fl\E\lt[(\Delta X_{s+t}-\Delta X_s)^2|\Lambda,D\rt]&=&4D\lt(1-\e^{-\Delta t\Lambda}-\e^{-t\Lambda}\lt(1-\f{\e^{-\Delta t\Lambda}+\e^{\Delta t\Lambda}}{2}\rt)\rt)\\
\fl&\xrightarrow{t\to\infty}&4D\lt(1-\e^{-\Delta t\Lambda}\rt).
\end{eqnarray}
The limit of the codifference is
\bgeq
\lim_{t\to\infty}\tau_{\Delta X}^\theta(t)=\f{1}{\theta^2}\ln\f{\E\lt[\e^{2\theta^2D\lt(\e^{-\Delta t\Lambda}-1\rt)}\rt]}{\E\lt[\e^{\theta^2D\lt(\e^{-\Delta t\Lambda}-1\rt)}\rt]^2}.
\eeq
Applying Jensen's inequality to the variable $\e^{2\theta^2D\e^{-\Delta t\Lambda}}$ and the function $x\mapsto x^2$ yields the inequality.

For the rescaled process it is straightforward to calculate that
\bgeq
\lim_{\Delta t\to 0^+}\lim_{t\to\infty}\tau_{\Delta \widetilde X}^\theta(t)=\lim_{\Delta t\to 0^+}\f{1}{\theta^2}\ln\f{\E\lt[\exp\lt(2\theta^2D\f{\e^{-\Delta t\Lambda}-1}{\Delta t}\rt)\rt]}{\E\lt[\exp\lt(\theta^2D\f{\e^{-\Delta t\Lambda}-1}{\Delta t}\rt)\rt]^2}= \f{1}{\theta^2}\ln\f{\E\lt[\e^{-2\theta^2D\Lambda}\rt]}{\E\lt[\e^{-\theta^2D\Lambda}\rt]^2}.
\eeq
\end{proof}

The last considered class of covariance functions is $Df(\Lambda)\exp(-t\Lambda)$. The increment process from Proposition \ref{prp:increments} fits this class with $f(\Lambda)=1-\exp(-\Delta t\Lambda)$, higher order increments and other similar transformations correspond to more complex $f$, but their behaviour at $0^+$ can be easily traced. Note that the proposition below is not a straightforward generalisation of Proposition \ref{prp:randLang}. The statements and methods of the derivation below are similar, but the assumptions do not coincide, because the introduction of the scaling $f(\Lambda)$ with a power law at $0$ was made at the cost of adding the strong requirement about the fast decay of tails of $D$, $\E[\exp(\theta^2D)]<\infty$:
\bg{prp}\label{prp:generalRandDL}
Let us consider the stationary, conditionally Gaussian process characterised by the conditional covariance
\bgeq
r_X(t|\Lambda,D) = D f(\Lambda)\e^{-t\Lambda}.
\eeq
Now, let us assume that $D$ and $\Lambda$ are independent, $\EE{\e^{\theta^2D}}<\infty$ and the PDF of $\Lambda$ has the form
\bgeq
p_\Lambda(\lambda)\sim L(\lambda)\lambda^{\alpha-1},\quad f(\lambda)\sim \lambda^\gamma, \quad \lambda\to 0^+; \quad \alpha,\gamma>0,
\eeq
for slowly varying function $L$. Then, for this class of processes
\bgeq\label{eq:langTauAs}
\tau_X^\theta(t)-\tau_X^\theta(\infty)\sim\f{\Gamma(\alpha+\gamma)\E[D]}{\EE{\e^{-\theta^2Df(\Lambda)}}}L(t^{-1}) t^{-\alpha-\gamma}.
\eeq
\end{prp}

\bg{proof}
We start from the formula
\bgeq
\tau_X^\theta(t)-\tau_X^\theta(\infty)=\f{1}{\theta^2}\ln\f{\E\lt[\e^{-\theta^2Df(\Lambda)\lt(1-\e^{-t\Lambda}\rt)}\rt]}{\E\lt[\e^{-\theta^2Df(\Lambda)}\rt]}
\eeq
which has the asymptotic
\begin{eqnarray}
\nonumber
\tau_X^\theta(t)-\tau_X^\theta(\infty)&\sim&\f{1}{\theta^2}\lt( \f{\E\lt[\e^{-\theta^2Df(\Lambda)\lt(1-\e^{-t\Lambda}\rt)}\rt]}{\E\lt[\e^{-\theta^2Df(\Lambda)}\rt]} - 1\rt)\\
&=&\f{1}{\theta^2\E\lt[\e^{-\theta^2Df(\Lambda)}\rt]}\E\lt[\e^{-\theta^2Df(\Lambda)}\lt(\e^{\theta^2Df(\Lambda)\e^{-t\Lambda}} -1\rt)\rt]
\end{eqnarray}
We thus need to study the tail behaviour of
\bgeq\label{eq:ISeries}
\E\lt[\e^{-\theta^2Df(\Lambda)}\sum_{k=1}^\infty\f{\theta^{2k}}{k!}D^kf(\Lambda)^k\e^{-kt\Lambda}\rt] = \sum_{k=1}^\infty\f{\theta^{2k}}{k!}\E\lt[D^k\E\lt[f(\Lambda)^k\e^{-\theta^2Df(\Lambda)}\e^{-kt\Lambda}|D\rt]\rt]
\eeq
We will analyse it using a bottom-up approach and start from considering the long time asymptotic of the conditional expected value $\E[\bd{\cdot}|D]$ for one term,
\begin{align}\label{eq:Df(L)}
&\f{t^{\alpha +k\gamma}}{L(1/t)} \E\lt[f(\Lambda)^k\e^{-\theta^2Df(\Lambda)}\e^{-kt\Lambda}|D\rt]= \f{t^{\alpha +k\gamma}}{L(1/t)}\int_0^\infty\dd\lambda\ f(\lambda)^k\e^{-\theta^2Df(\lambda)}p_\Lambda(\lambda)\e^{-kt\lambda} \nonumber\\
&=\f{t^{\alpha +k\gamma-1}}{L(1/t)}\f{1}{k}\int_0^\infty\dd\lambda\ f\lt(\f{\lambda}{kt}\rt)^k\e^{-\theta^2Df\lt(\f{\lambda}{kt}\rt)}p_\Lambda\lt(\f{\lambda}{kt}\rt)\e^{-\lambda} \nonumber\\
&=\f{1}{k^{\alpha+k\gamma}}\int_0^\infty\dd\lambda\ \lt(\f{f\lt(\f{\lambda}{kt}\rt)}{\lt(\f{\lambda}{kt}\rt)^\gamma}\rt)^k \e^{-\theta^2Df\lt(\f{\lambda}{kt}\rt)}\f{p_\Lambda\lt(\f{\lambda}{kt}\rt)}{L(1/t)\lt(\f{\lambda}{kt}\rt)^{\alpha-1}} \lambda^{\alpha+k\gamma-1}\e^{-\lambda} \nonumber\\
&\xrightarrow{t\to\infty} \f{1}{k^{\alpha+k\gamma}}\int_0^\infty\dd\lambda\ \lambda^{\alpha+k\gamma-1}\e^{-\lambda} = \f{\Gamma(\alpha+k\gamma)}{k^{\alpha+k\gamma}}.
\end{align}
Now  take $\delta>0$ such that $f(\lambda)<1$ for all $0\le \lambda<\delta $ and $\epsilon>0$ such that $L(1/t)>t^{-1/2}$ for sufficiently large $t$

\begin{align}
&\f{D^k}{k!}\f{t^{\alpha +\gamma}}{L(1/t)}f(\Lambda)^k\e^{-\theta^2Df(\Lambda)}\e^{-kt\Lambda}
\le \f{D^k}{k!}\f{t^{\alpha +\gamma}}{L(1/t)}\lt( f(\Lambda)\e^{-t\Lambda}\bd{1}_{\Lambda<\delta}+f(\Lambda)^k\e^{-\theta^2D f(\Lambda)}\e^{-kt\delta}\bd{1}_{\Lambda\ge \delta}\rt)\nonumber\\
&\le \f{D^k}{k!}\f{t^{\alpha +\gamma}}{L(1/t)}
f(\Lambda)\e^{-t\Lambda}+ \f{k^k}{k!}\e^{-k}t^{\alpha +\gamma+1/2}\e^{-k t\delta},
\end{align}
where we additionally used the inequality $x^k\e^{-x}\le k^k\e^{-k}$.
Now, for the left term above observe that
\bgeq
\f{t^{\alpha +\gamma}}{L(1/t)}
\E\lt[f(\Lambda)\e^{-t\Lambda}\rt]\xrightarrow{t\to\infty}\Gamma(\alpha),
\eeq
so it is bounded with respect to $t$ by some constant, let it be $c_1$,
\bgeq
\sum_{k=1}^\infty \E\lt[\f{\theta^{2k}D^k}{k!}\f{t^{\alpha +\gamma}}{L(1/t)}
f(\Lambda)\e^{-t\Lambda}\rt]\le c_1\sum_{k=1}^\infty \f{\theta^{2k}\E\lt[D^k\rt]}{k!} = c_1\E\lt[\e^{\theta^2D}\rt].
\eeq
And for the right term, the Stirling formula shows that
\bgeq
\f{k^k}{k!}\e^{-k}\sim \sqrt{2\pi}k^{-1/2},\quad k\to\infty.
\eeq
Moreover straightforward calculation yields
\bgeq
t^{\alpha+\gamma+1/2}\e^{-kt\delta}\le c_2 k^{-\alpha-\gamma-1/2},
\eeq
so the whole series behaves like $k^{-1-\alpha-\gamma}$ and is summable.

Therefore, we have shown that we can use the dominated convergence theorem with respect to series \Ref{eq:ISeries} multiplied by $t^{\alpha+\gamma}/L(1/t)$. According to \Ref{eq:Df(L)} the term $k=1$ converges to $\E[D]\Gamma(\alpha+\gamma)$ and all terms $k>1$ decay like $t^{-k\gamma}$. Only the first term remains in the limit $t\to\infty$ and
\bgeq
\f{t^{\alpha+\gamma}}{L(1/t)}\sum_{k=1}^\infty\f{\theta^{2k}}{k!}\E\lt[D^k\E\lt[f(\Lambda)^k\e^{-\theta^2Df(\Lambda)}\e^{-kt\Lambda}|D\rt]\rt]\xrightarrow{t\to\infty} \theta^2\E[D]\Gamma(\alpha+\gamma).
\eeq
\end{proof}

\noindent\textbf{Remark.} Propositions \ref{prp:randLang} and \ref{prp:generalRandDL} above can be generalised by replacing $t$ by $g(t)$ in the formula for the covariance, the only requirement is that $g(t)\to\infty$ as $t\to\infty$. This allows one to consider some more general types of the dependence, e.g., the power-law $t^{-2H}$ corresponds to $\Lambda=2H$ and $g(t)=\ln(t)$.

\subsection{Diffusing diffusivity}
\bg{prp}\label{prp:prod}

Let us assume that $Y$ and $Z$ are centred stationary Gaussian processes. Without loss of generality we assume $\E[Y_t^2]=\E[Z_t^2]=1$. Let $X$ be given by 
\bg{itemize}
\item[a)]
\bgeq
X_t = (\sigma Z_t+d)Y_t,
\eeq
\item[b)]
\bgeq
X_t=\sigma |Z_t|Y_t;
\eeq
\end{itemize}
with deterministic $\sigma,d>0$. Then the codifference of $X$ is given by elementary formulae, as given at the end of corresponding derivations in Eqs. \Ref{eq:ZYa} and \Ref{eq:ZYb}.
\end{prp}

\bg{proof}
We begin by conditioning over $Z_t$, the averages then become Gaussian averages rescaled by values $Z_t$. Next we calculate the denominator in the codifference
\bgeq
\E\lt[\e^{\I\theta X_{s+t}}\rt]\E\lt[\e^{-\I\theta X_s}\rt] = \E\lt[\e^{-\theta^2(\sigma Z_s+d)^2/2}\rt]^2= \f{1}{1+(\theta\sigma)^2}\exp\lt(-\theta^2\f{d^2}{1+(\theta\sigma)^2}\rt).
\eeq
The last equality corresponds to calculating a Gaussian integral, which can also be interpreted as a Laplace transform of the distribution $\chi^2(1)$. The numerator is more complicated,
\bgeq\label{eq:Zexp}
\E\lt[\e^{\I\theta(X_{s+t}-X_s)}\rt]= \E\lt[\e^{-\theta^2\lt((\sigma Z_{s+t}+d)^2+(\sigma Z_s+d)^2-2r_Y(t)(\sigma Z_{s+t}+d)(\sigma Z_s+d)\rt)/2}\rt].
\eeq
The above expectation can be calculated if we decompose $[Z_{s+t},Z_s]\deq [A_++A_-,A_+-A_-]$ where $A_+,A_-$ are independent Gaussian variables, whose variances can be found to be $\E[A_\pm^2]=(1\pm r_Z(t))/2$. After substitution the exponent in \Ref{eq:Zexp} factorises into
\begin{eqnarray}
\nonumber
\E\lt[\e^{\I\theta (X_{s+t}-X_s)}\rt]&=&\E\lt[\e^{-(\theta\sigma)^2(1-r_Y(t))A_+^2-2(\theta\sigma)^2(1-r_Y(t))A_+d}\rt]\\
&&\times\E\lt[\e^{-(\theta\sigma)^2(1+r_Y(t))A_-^2}\rt] \e^{-\theta^2(1-r_Y(t))d^2}.
\end{eqnarray}
Both obtained terms are Gaussian integrals which can be easily evaluated. Taking both together and calculating the logarithm we obtain
\bg{align}\label{eq:ZYa}
\tau_X^\theta(t) &= \f{(1-r_Y(t))^2((\theta\sigma)^2+r_Z(t))}{1+(1-r_Y(t))((\theta\sigma)^2+r_Z(t))}d^2+\f{d^2}{1+(\theta\sigma)^2}-(1-r_Y(t))d^2+\f{1}{\theta^2}\ln(1+(\theta\sigma)^2)\nonumber\\
&-\f{1}{2\theta^2} \ln\Big(\big(1+((\theta\sigma)^2+r_Z(t))(1-r_Y(t))\big)\big(1+((\theta\sigma)^2-r_Z(t))(1+r_Y(t))\big)\Big).
\end{align}

For b) the denominator is simple and yields $(1+(\theta\sigma)^2)^{-1}$. The numerator can be expressed as
\bgeq
\EE{\e^{\I\theta\sigma(|Z_{s+t}|Y_{s+t}-|Z_s|Y_s})} = \EE{\exp\lt(-\f{(\theta\sigma)^2}{2}\lt(Z_1^2+Z_2^2-2r_Y(t)|Z_1Z_2|\rt)\rt)}.
\eeq
Using the formula for the two-dimensional density of $[Z_{s+t},Z_s]$, the term under the logarithm in the formula for the codifference can be expressed as an integral over the function  
\bg{align}\label{eq:funz1z2}
& C\exp\lt(-\f{(\theta\sigma)^2}{2}\big(z_1^2+z_2^2-2r_Y(t)|z_1z_2|\big)\rt)\exp\lt(-\f{1}{2(1-r_Z(t)^2)}\lt( z_1^2+z_2^2-2r_Z(t)z_1z_1\rt)\rt)\nonumber\\
& =C\exp\lt(-sz_1^2-sz_2^2+2 s \rho_{\mathrm{sgn}(z_1z_2)}|z_1z_2|      \rt),
\end{align}
where
\begin{eqnarray}
\nonumber
&& C \defeq \f{1+(\theta\sigma)^2}{2\pi\sqrt{1-r_Z(t)^2}},\quad s \defeq \f{1}{2}\lt((\theta\sigma)^2+\f{1}{1-r_Z(t)^2}\rt),\\
&&\rho_\pm \defeq \f{ (\theta\sigma)^2(1-r_Z(t)^2)r_Y(t)\pm r_Z(t)}{(\theta\sigma)^2(1-r_Z(t)^2)+1}.
\end{eqnarray}

The integration over $\R^2$ of  \Ref{eq:funz1z2} can be changed to an integration over $\R_+^2$ 
\bg{align}\label{eq:ZYb}
&I = 2C\int_{0}^\infty\dd z_1\!\int_0^\infty\dd z_2\ \e^{-sz_1^2-sz_2^2}\sum_{\rho\in\{\rho_+,\rho_-\}}\e^{2s\rho z_1z_2}\nonumber\\
&=\f{2C}{s}\int_{0}^\infty\dd z_1\!\int_0^\infty\dd z_2\ \e^{-z_1^2-z_2^2} \sum_{\rho\in\{\rho_+,\rho_-\}}\e^{2\rho z_1z_2}\nonumber\\
&= \f{\sqrt{\pi}C}{s}\sum_{\rho\in\{\rho_+,\rho_-\}}\int_0^\infty\dd z_1\ \e^{-(1-\rho)z_1^2}\mathrm{erfc}\lt(-\rho z_1\rt)\nonumber\\
& = \f{\sqrt{1-r_Z(t)^2}}{\pi}\f{1+(\theta\sigma)^2}{(\theta\sigma)^2(1-r_Z(t)^2)+1}\sum_{\rho\in\{\rho_+,\rho_-\}}\f{\f{\pi}{2}+\mathrm{arctan}\lt(\f{\rho}{\sqrt{1-\rho^2}}\rt)}{\sqrt{1-\rho^2}}.
\end{align}
Now, the codifference is  $\tau_X^\theta(t)=\theta^{-2}\ln I$.
\end{proof}
When $r_Y(t)=0$ the above formulae simplify significantly and simple asymptotic can be derived by direct computation, see Eqs. \Ref{eq:diffDiffAs1} and \Ref{eq:diffDiffAs2}. The case $r_Z(t)=0$ also leads to a simplification and can be considered in a more general setting. 

\bg{prp}\label{prp:indepD}
If $Y_t$ is a stationary Gaussian process, $\E[Y_t^2]=1$ and for large enough $t$ values $D_s$ and $D_{s+t}$ are i.i.d and independent of $Y$, then for $X_t=\sqrt{D_t}Y_t$
\bgeq
\tau_X^\theta(t)\sim \f{\EE{\sqrt{D}\e^{-\theta^2D/2}}^2}{\EE{\e^{-\theta^2D/2}}^2}r_Y(t),\quad r_Y(t)\to 0,
\eeq
where $D$ has the same distribution as $D_s$ or $D_{s+t}$.
\end{prp}
\bg{proof}
We take $t$ large enough so that we can represent the values of $X$ as $X_s = \sqrt{D_1}Y_s$ and $X_{s+t}=\sqrt{D_2}Y_{s+t}$ for i.i.d. $D_1$ and $D_2$. Using a conditioning on $D_1,D_2$ the codifference can be expressed as
\begin{eqnarray}
\nonumber
\tau_X^\theta(t)&=&\f{1}{\theta^2}\ln \f{\EE{\e^{-\theta^2\big(D_1+D_2-2r_Y(t)\sqrt{D_1D_2}\big)/2}}}{\EE{\e^{-\theta^2(D_1+D_2)/2}}}\\
&=&\f{1}{\theta^2}\ln\lt( \f{\EE{\e^{-\theta^2(D_1+D_2)/2}\lt(\e^{\theta^2r_Y(t)\sqrt{D_1D_2}}-1\rt)}}{\EE{\e^{-\theta^2(D_1+D_2)/2}}}+1\rt).
\end{eqnarray}
Now we consider the numerator in the above, divide it by $r_Y(t)$ and, using dominated convergence as in previous propositions,
\bgeq
\EE{\e^{-\theta^2(D_1+D_2)/2}\f{\e^{\theta^2r_Y(t)\sqrt{D_1D_2}}-1}{r_Y(t)}}\xrightarrow{r_Y(t)\to 0}\theta^2 \EE{\sqrt{D_1D_2}\e^{-\theta^2(D_1+D_2)/2}}.
\eeq
The result follows.
\end{proof}

\appendix

\section{Sample size dependence of codifference and covariance}

In supplement to figure \ref{fig:MC} we show in figure \ref{addfig} that even
for smaller sample sizes such as $10^4$, $10^3$, and 500 significant differences
between the covariance and codifference of increments are visible.

\begin{figure}
\includegraphics[width=0.3\textwidth]{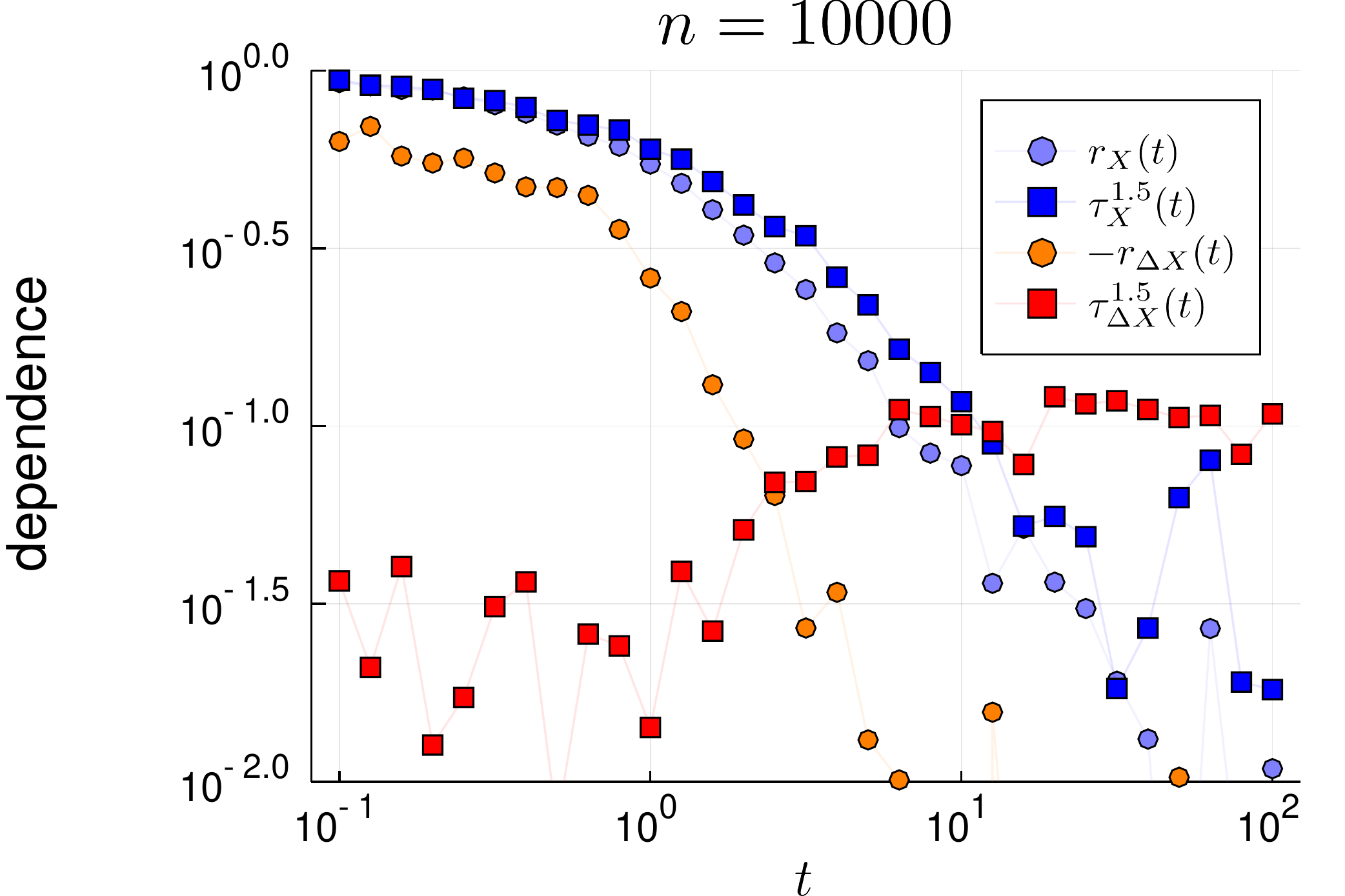}
\includegraphics[width=0.3\textwidth]{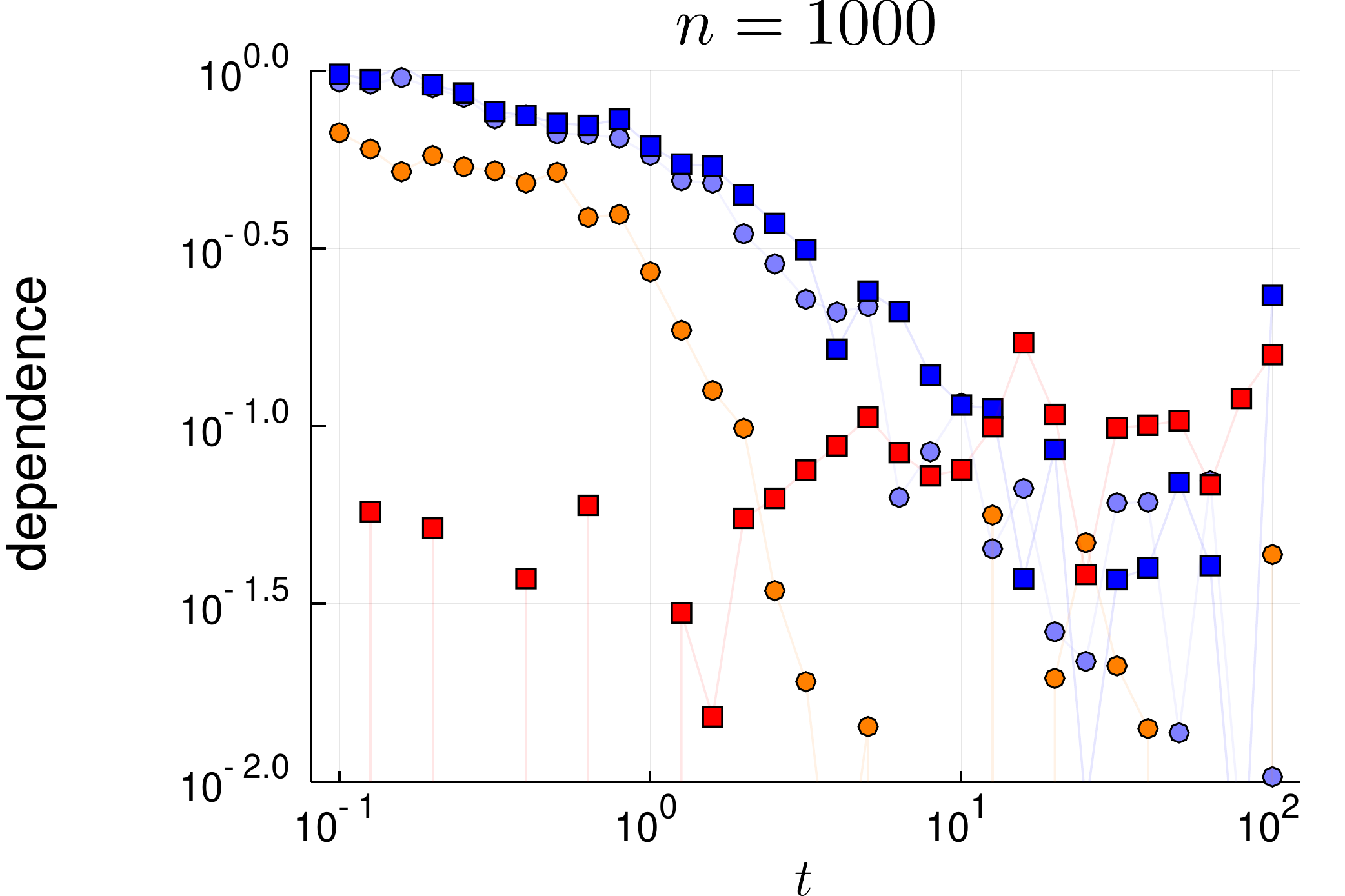}
\includegraphics[width=0.3\textwidth]{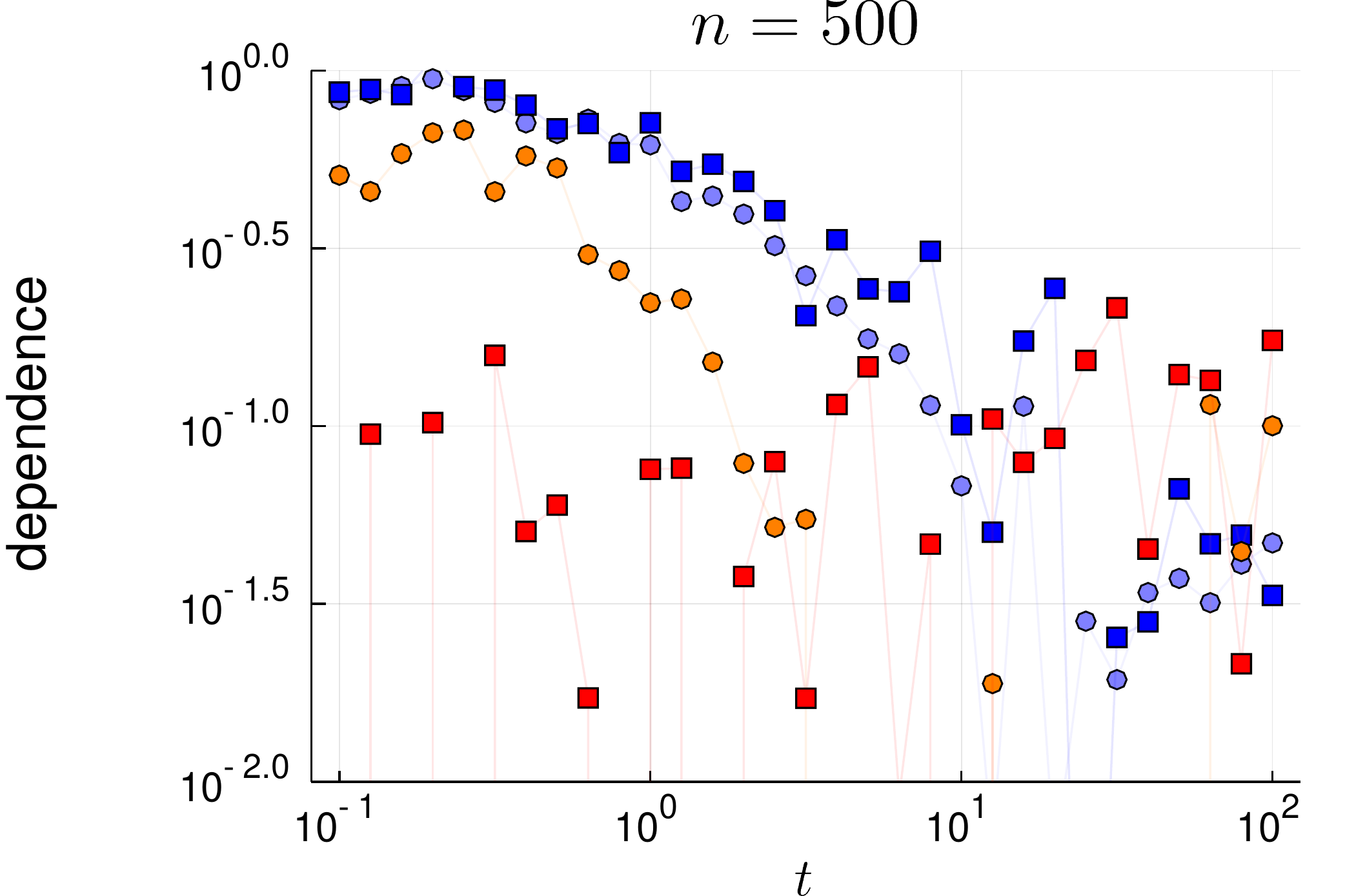}
\caption{Estimated codifference $\tau$ and covariance $r$ for
sample sizes (from left to right) $10^4$, $10^3$, and 500.}
\label{addfig}
\end{figure}

\ack

We acknoeldge funding from the Polish National Science Centre, HARMONIA 8 grant no.
UMO-2016/22/M/ST1/00233, and from Deutsche Forschungsgemeinschaft, grants ME1535/6-1
and ME1535/7-1. RM was supported by an Alexander von Humboldt Polish Honorary
Research Scholarship from the Foundation for Polish Science (Fundacja na rzecz
Nauki Polski).

\end{document}